\documentclass[
amsmath,
amssymb,
prd,
twocolumn,
superscriptaddress,
floatfix,
]{revtex4}

\usepackage{ifthen}
\usepackage{epsfig}
\usepackage{amsmath}
\usepackage{comment}
\usepackage{longtable}
\usepackage{subfigure}
\usepackage{color}
\usepackage{xspace}
\usepackage{slashbox}
\usepackage{amsfonts}

\usepackage{multirow}
\usepackage{array} 
\usepackage{booktabs}
\usepackage{dcolumn}
\usepackage[abs]{overpic}

\def\simge{\mathrel{%
   \rlap{\raise 0.511ex \hbox{$>$}}{\lower 0.511ex \hbox{$\sim$}}}}
\def\simle{\mathrel{
   \rlap{\raise 0.511ex \hbox{$<$}}{\lower 0.511ex \hbox{$\sim$}}}}
\newcommand{\WW}{{\it WW}}
\newcommand{\WZ}{{\it WZ}}
\newcommand{\ZZ}{{\it ZZ}}
\newcommand{\nunubar}{\nu \overline{\nu}}
\newcommand{\llll}{\ell^+\ell^-\ell^{^{(_{\null'})}+}\ell^{^{(_{\null'})}-}}

\newcommand{\llvv}{\ell^+\ell^-\nunubar}
\newcommand{\ppbar}{p \overline{p}}
\newcommand{\ttbar}{t \overline{t}}

\newcommand{\fb}{\rm{fb}^{-1}}
\newcommand{\pt}{p_T}
\newcommand{\Et}{E_T}
\newcommand{\MET}{\mbox{$E\kern-0.50em\raise0.10ex\hbox{/}_{T}$}}
\newcommand{\MPT}{\mbox{$p\kern-0.50em\raise0.10ex\hbox{/}_{T}$}}

\begin{document}
\title{Measurement of the \ZZ~production cross section using the 
full CDF II data set}
\affiliation{Institute of Physics, Academia Sinica, Taipei, Taiwan 11529, Republic of China}
\affiliation{Argonne National Laboratory, Argonne, Illinois 60439, USA}
\affiliation{University of Athens, 157 71 Athens, Greece}
\affiliation{Institut de Fisica d'Altes Energies, ICREA, Universitat Autonoma de Barcelona, E-08193, Bellaterra (Barcelona), Spain}
\affiliation{Baylor University, Waco, Texas 76798, USA}
\affiliation{Istituto Nazionale di Fisica Nucleare Bologna, \ensuremath{^{ii}}University of Bologna, I-40127 Bologna, Italy}
\affiliation{University of California, Davis, Davis, California 95616, USA}
\affiliation{University of California, Los Angeles, Los Angeles, California 90024, USA}
\affiliation{Instituto de Fisica de Cantabria, CSIC-University of Cantabria, 39005 Santander, Spain}
\affiliation{Carnegie Mellon University, Pittsburgh, Pennsylvania 15213, USA}
\affiliation{Enrico Fermi Institute, University of Chicago, Chicago, Illinois 60637, USA}
\affiliation{Comenius University, 842 48 Bratislava, Slovakia; Institute of Experimental Physics, 040 01 Kosice, Slovakia}
\affiliation{Joint Institute for Nuclear Research, RU-141980 Dubna, Russia}
\affiliation{Duke University, Durham, North Carolina 27708, USA}
\affiliation{Fermi National Accelerator Laboratory, Batavia, Illinois 60510, USA}
\affiliation{University of Florida, Gainesville, Florida 32611, USA}
\affiliation{Laboratori Nazionali di Frascati, Istituto Nazionale di Fisica Nucleare, I-00044 Frascati, Italy}
\affiliation{University of Geneva, CH-1211 Geneva 4, Switzerland}
\affiliation{Glasgow University, Glasgow G12 8QQ, United Kingdom}
\affiliation{Harvard University, Cambridge, Massachusetts 02138, USA}
\affiliation{Division of High Energy Physics, Department of Physics, University of Helsinki, FIN-00014, Helsinki, Finland; Helsinki Institute of Physics, FIN-00014, Helsinki, Finland}
\affiliation{University of Illinois, Urbana, Illinois 61801, USA}
\affiliation{The Johns Hopkins University, Baltimore, Maryland 21218, USA}
\affiliation{Institut f\"{u}r Experimentelle Kernphysik, Karlsruhe Institute of Technology, D-76131 Karlsruhe, Germany}
\affiliation{Center for High Energy Physics: Kyungpook National University, Daegu 702-701, Korea; Seoul National University, Seoul 151-742, Korea; Sungkyunkwan University, Suwon 440-746, Korea; Korea Institute of Science and Technology Information, Daejeon 305-806, Korea; Chonnam National University, Gwangju 500-757, Korea; Chonbuk National University, Jeonju 561-756, Korea; Ewha Womans University, Seoul, 120-750, Korea}
\affiliation{Ernest Orlando Lawrence Berkeley National Laboratory, Berkeley, California 94720, USA}
\affiliation{University of Liverpool, Liverpool L69 7ZE, United Kingdom}
\affiliation{University College London, London WC1E 6BT, United Kingdom}
\affiliation{Centro de Investigaciones Energeticas Medioambientales y Tecnologicas, E-28040 Madrid, Spain}
\affiliation{Massachusetts Institute of Technology, Cambridge, Massachusetts 02139, USA}
\affiliation{University of Michigan, Ann Arbor, Michigan 48109, USA}
\affiliation{Michigan State University, East Lansing, Michigan 48824, USA}
\affiliation{Institution for Theoretical and Experimental Physics, ITEP, Moscow 117259, Russia}
\affiliation{University of New Mexico, Albuquerque, New Mexico 87131, USA}
\affiliation{The Ohio State University, Columbus, Ohio 43210, USA}
\affiliation{Okayama University, Okayama 700-8530, Japan}
\affiliation{Osaka City University, Osaka 558-8585, Japan}
\affiliation{University of Oxford, Oxford OX1 3RH, United Kingdom}
\affiliation{Istituto Nazionale di Fisica Nucleare, Sezione di Padova, \ensuremath{^{jj}}University of Padova, I-35131 Padova, Italy}
\affiliation{University of Pennsylvania, Philadelphia, Pennsylvania 19104, USA}
\affiliation{Istituto Nazionale di Fisica Nucleare Pisa, \ensuremath{^{kk}}University of Pisa, \ensuremath{^{ll}}University of Siena, \ensuremath{^{mm}}Scuola Normale Superiore, I-56127 Pisa, Italy, \ensuremath{^{nn}}INFN Pavia, I-27100 Pavia, Italy, \ensuremath{^{oo}}University of Pavia, I-27100 Pavia, Italy}
\affiliation{University of Pittsburgh, Pittsburgh, Pennsylvania 15260, USA}
\affiliation{Purdue University, West Lafayette, Indiana 47907, USA}
\affiliation{University of Rochester, Rochester, New York 14627, USA}
\affiliation{The Rockefeller University, New York, New York 10065, USA}
\affiliation{Istituto Nazionale di Fisica Nucleare, Sezione di Roma 1, \ensuremath{^{pp}}Sapienza Universit\`{a} di Roma, I-00185 Roma, Italy}
\affiliation{Mitchell Institute for Fundamental Physics and Astronomy, Texas A\&M University, College Station, Texas 77843, USA}
\affiliation{Istituto Nazionale di Fisica Nucleare Trieste, \ensuremath{^{qq}}Gruppo Collegato di Udine, \ensuremath{^{rr}}University of Udine, I-33100 Udine, Italy, \ensuremath{^{ss}}University of Trieste, I-34127 Trieste, Italy}
\affiliation{University of Tsukuba, Tsukuba, Ibaraki 305, Japan}
\affiliation{Tufts University, Medford, Massachusetts 02155, USA}
\affiliation{University of Virginia, Charlottesville, Virginia 22906, USA}
\affiliation{Waseda University, Tokyo 169, Japan}
\affiliation{Wayne State University, Detroit, Michigan 48201, USA}
\affiliation{University of Wisconsin, Madison, Wisconsin 53706, USA}
\affiliation{Yale University, New Haven, Connecticut 06520, USA}

\author{T.~Aaltonen}
\affiliation{Division of High Energy Physics, Department of Physics, University of Helsinki, FIN-00014, Helsinki, Finland; Helsinki Institute of Physics, FIN-00014, Helsinki, Finland}
\author{S.~Amerio\ensuremath{^{jj}}}
\affiliation{Istituto Nazionale di Fisica Nucleare, Sezione di Padova, \ensuremath{^{jj}}University of Padova, I-35131 Padova, Italy}
\author{D.~Amidei}
\affiliation{University of Michigan, Ann Arbor, Michigan 48109, USA}
\author{A.~Anastassov\ensuremath{^{v}}}
\affiliation{Fermi National Accelerator Laboratory, Batavia, Illinois 60510, USA}
\author{A.~Annovi}
\affiliation{Laboratori Nazionali di Frascati, Istituto Nazionale di Fisica Nucleare, I-00044 Frascati, Italy}
\author{J.~Antos}
\affiliation{Comenius University, 842 48 Bratislava, Slovakia; Institute of Experimental Physics, 040 01 Kosice, Slovakia}
\author{G.~Apollinari}
\affiliation{Fermi National Accelerator Laboratory, Batavia, Illinois 60510, USA}
\author{J.A.~Appel}
\affiliation{Fermi National Accelerator Laboratory, Batavia, Illinois 60510, USA}
\author{T.~Arisawa}
\affiliation{Waseda University, Tokyo 169, Japan}
\author{A.~Artikov}
\affiliation{Joint Institute for Nuclear Research, RU-141980 Dubna, Russia}
% ...

\collaboration{CDF Collaboration}
\altaffiliation[With visitors from]{
\ensuremath{^{a}}University of British Columbia, Vancouver, BC V6T 1Z1, Canada,
\ensuremath{^{b}}Istituto Nazionale di Fisica Nucleare, Sezione di Cagliari, 09042 Monserrato (Cagliari), Italy,
\ensuremath{^{c}}University of California Irvine, Irvine, CA 92697, USA,
\ensuremath{^{d}}Institute of Physics, Academy of Sciences of the Czech Republic, 182~21, Czech Republic,
\ensuremath{^{e}}CERN, CH-1211 Geneva, Switzerland,
\ensuremath{^{f}}Cornell University, Ithaca, NY 14853, USA,
\ensuremath{^{g}}University of Cyprus, Nicosia CY-1678, Cyprus,
\ensuremath{^{h}}Office of Science, U.S. Department of Energy, Washington, DC 20585, USA,
\ensuremath{^{i}}University College Dublin, Dublin 4, Ireland,
\ensuremath{^{j}}ETH, 8092 Z\"{u}rich, Switzerland,
\ensuremath{^{k}}University of Fukui, Fukui City, Fukui Prefecture, Japan 910-0017,
\ensuremath{^{l}}Universidad Iberoamericana, Lomas de Santa Fe, M\'{e}xico, C.P. 01219, Distrito Federal,
\ensuremath{^{m}}University of Iowa, Iowa City, IA 52242, USA,
\ensuremath{^{n}}Kinki University, Higashi-Osaka City, Japan 577-8502,
\ensuremath{^{o}}Kansas State University, Manhattan, KS 66506, USA,
\ensuremath{^{p}}Brookhaven National Laboratory, Upton, NY 11973, USA,
\ensuremath{^{q}}Queen Mary, University of London, London, E1 4NS, United Kingdom,
\ensuremath{^{r}}University of Melbourne, Victoria 3010, Australia,
\ensuremath{^{s}}Muons, Inc., Batavia, IL 60510, USA,
\ensuremath{^{t}}Nagasaki Institute of Applied Science, Nagasaki 851-0193, Japan,
\ensuremath{^{u}}National Research Nuclear University, Moscow 115409, Russia,
\ensuremath{^{v}}Northwestern University, Evanston, IL 60208, USA,
\ensuremath{^{w}}University of Notre Dame, Notre Dame, IN 46556, USA,
\ensuremath{^{x}}Universidad de Oviedo, E-33007 Oviedo, Spain,
\ensuremath{^{y}}CNRS-IN2P3, Paris, F-75205 France,
\ensuremath{^{z}}Universidad Tecnica Federico Santa Maria, 110v Valparaiso, Chile,
\ensuremath{^{aa}}The University of Jordan, Amman 11942, Jordan,
\ensuremath{^{bb}}Universite catholique de Louvain, 1348 Louvain-La-Neuve, Belgium,
\ensuremath{^{cc}}University of Z\"{u}rich, 8006 Z\"{u}rich, Switzerland,
\ensuremath{^{dd}}Massachusetts General Hospital, Boston, MA 02114 USA,
\ensuremath{^{ee}}Harvard Medical School, Boston, MA 02114 USA,
\ensuremath{^{ff}}Hampton University, Hampton, VA 23668, USA,
\ensuremath{^{gg}}Los Alamos National Laboratory, Los Alamos, NM 87544, USA,
\ensuremath{^{hh}}Universit\`{a} degli Studi di Napoli Federico I, I-80138 Napoli, Italy
}
\noaffiliation

\begin{abstract}
We present a measurement of the \ZZ~boson-pair production cross section 
in 1.96~TeV center-of-mass energy $\ppbar$ collisions. We reconstruct 
final states incorporating four charged leptons or two charged leptons 
and two neutrinos from the full data set collected by the Collider 
Detector experiment at the Fermilab Tevatron, corresponding to 9.7~$\fb$ 
of integrated luminosity.  Combining the results obtained from each final 
state, we measure a cross section of 1.04$^{+0.32}_{-0.25}$~pb, in good 
agreement with the standard model prediction at next-to-leading order 
in the strong-interaction coupling.
\end{abstract}

\pacs{PACS}

\maketitle

\section{Introduction \label{sec:intro}}

Measurements of production cross sections for electroweak vector boson 
pairs are important tests of standard model (SM) predictions in the 
electroweak sector. If sufficiently precise, these measurements may 
signal contributions from non-SM physics such as anomalous trilinear 
gauge couplings~\cite{trilinear} or large extra dimensions~\cite{extraD}, 
which can enhance or suppress diboson production rates.

At next-to-leading order (NLO) accuracy in the strong coupling constant 
$\alpha_s$, the SM prediction of the \ZZ~production cross section 
for $\ppbar$ collisions at a center-of-mass energy of 1.96~TeV is 
$1.4~\pm~0.1$~pb~\cite{SMXS-old}, which is among the smallest cross 
sections accessible by the Tevatron experiments.

The \ZZ~production process was first studied at the LEP $e^+e^-$ collider 
at CERN~\cite{ZZ_aleph,ZZ_delphi,ZZ_l3,ZZ_opal} and more recently at 
the Tevatron $\ppbar$ collider. A previous \ZZ~production cross section 
measurement from leptonic final states has been published by the CDF 
collaboration using 6~$\fb$ of integrated luminosity~\cite{CDF_ZZ_59}, 
the result of which is extended here.  The most recent D0 collaboration 
results from the four charged lepton~\cite{D0_HZZ_full} and two charged 
lepton plus two neutrino~\cite{D0_llvv_pub} final states, using up to 9.8~$\fb$ 
and 8.6~$\fb$ of data respectively, have been combined~\cite{D0_HZZ_full} 
to obtain a \ZZ~production cross section measurement of 
$1.32^{+0.32}_{-0.28}$~pb.  Further studies have been performed at the 
LHC $pp$ collider at CERN, where the ATLAS and CMS experiments have 
carried out measurements using data collected through the 2012 collider 
run~\cite{ZZ_atlas_2012,ZZ_atlas_2013,ZZ_cms_2013,ZZ_cms_2013_2}.

In this article we present a measurement of the \ZZ~production 
cross section using the full data sample collected by the CDF~II 
detector~\cite{CDFIIdet} at the Tevatron, corresponding to 9.7~$\fb$ 
of integrated luminosity. Compared to previous CDF studies, we 
analyze the full data sample and optimize the event selection further 
to reduce background contributions and obtain an improved measurement 
accuracy. The cross section is independently measured from two leptonic 
final states, $\llll$ and $\llvv$, where $\ell$ and $\ell^{^{(_{\null'})}}$ 
indicate electrons or muons originating either from prompt decays of 
the $Z$ boson or from leptonic decays of a $\tau$ lepton in cases where 
the $Z$ boson decays into a $\tau$ lepton pair.  The portions of the 
inclusive $\ppbar\to Z/\gamma^* Z/\gamma^*$ cross section accessible 
to the $\llll$ and $\llvv$ decay modes are somewhat different due to 
the absence of $\gamma^*$ couplings to neutrinos.  In order to combine 
the measurements, both are extrapolated to an inclusive \ZZ~production 
cross section assuming the zero-width approximation, where the 
contributions from $\gamma^*\gamma^*$ production and $Z\gamma^*$ 
interference are set to zero.  Henceforth, we use \ZZ~to denote the 
inclusive $Z/\gamma^* Z/\gamma^*$ production process.

The article is structured as follows. Section~\ref{sec:cdf} contains 
a brief description of the CDF~II detector. Section~\ref{sec:event} 
discusses how \ZZ~events are identified in the detector. 
Sections~\ref{sec:4leps} and~\ref{sec:llvv} describe the measurement 
techniques used for the $\llll$  and $\llvv$ final states, respectively. 
The combination of the two measurements used to obtain the final 
production cross section result is described in Section~\ref{sec:combo}.

\section{The CDF~II detector \label{sec:cdf}}

The components of the CDF~II detector relevant to this analysis are 
briefly described below, while a complete description can be found 
elsewhere~\cite{CDFIIdet}. The detector geometry is described using 
the azimuthal angle $\phi$ and the pseudorapidity $\eta\equiv -\ln 
(\tan \theta/2 )$, where $\theta$ is the polar angle of a particle's 
trajectory with respect to the proton beam axis (positive $z$-axis). 
The pseudorapidity relative to the center of the detector is referred 
to as $\eta_{\mathrm{det}}$.  Transverse energy and momentum are 
defined as $E_T\equiv E\sin\theta$ and $p_T\equiv p\sin\theta$, 
where $E$ is the energy measured in a calorimeter tower (or related 
to an energy cluster) and $p$ is a charged particle momentum.  
The trajectories of charged particles (tracks) are reconstructed 
using silicon micro-strip detectors~\cite{SVXII} and a 96-layer 
open-cell drift chamber (COT)~\cite{COT} located in a 1.4~T solenoidal 
magnetic field. The plateau of the drift chamber acceptance covers 
$|\eta_{\mathrm{det}}|\leq$~1. The inner silicon tracker provides 
coverage of up to 8 layers with radii between 1.35~cm and 28~cm in 
the region $|\eta_{\mathrm{det}}|\leq$~2. Electromagnetic (EM) and 
hadronic (HAD) sampling calorimeters segmented in a projective-tower 
geometry are located outside the solenoid. At depths corresponding 
to one hadronic-interaction length ($\lambda$), which is equivalent 
to 18--20 radiation lengths ($X_0$), lead absorber is used to 
measure the electromagnetic component of showers, while in the 
region 4.5--7~$\lambda$ iron is used to contain the hadronic 
component.  A central calorimeter covers the pseudorapidity region  
$|\eta_{\mathrm{det}}|\leq$~1.1, and a forward calorimeter extends 
the coverage into the region 1.1~$\leq |\eta_{\mathrm{det}}|\leq$~3.6.
Shower-maximum detectors (SMX) embedded in the electromagnetic 
calorimeters at a depth approximately corresponding to 6~$X_0$ assist 
in the position measurement and background suppression for electrons. 
Drift chambers and scintillators are located outside the calorimeter 
to identify muons, which approximate minimum-ionizing particles and 
typically deposit only a fraction of their energy in the absorber 
material. 

\section{Event reconstruction \label{sec:event}}

We collect \ZZ~candidate events using an online event-selection system 
(trigger) that records events satisfying at least one of several 
high-$\pt$ lepton requirements. The central electron trigger requires 
an EM energy deposit (clustered among towers in the calorimeter) with 
$E_T \geq$~18~GeV matched to a charged particle with $p_T \geq$~8~GeV/$c$. 
Several muon triggers based on muon drift-chamber track segments 
({\it stubs}) matched to charged particles with $\pt\geq$~18~GeV/$c$ 
are also incorporated. Trigger selection efficiencies are measured 
from collected event samples containing leptonic $W$ and $Z$ boson 
decays~\cite{trig_eff}.

We use three complementary track pattern recognition algorithms 
distinguished by their starting point: hits in the COT, hits in the 
silicon tracker, or the projections of observed calorimeter energy 
clusters back to the interaction region (calorimeter-seeded tracks).
Electrons are identified by matching a reconstructed track to an 
energy cluster reconstructed in the EM calorimeters. Muons are 
identified by matching a track to an energy deposit in the calorimeter 
consistent with originating from a minimum-ionizing particle, with 
or without an associated stub in the muon system.  All leptons are 
required to be isolated such that the sum of additional $\Et$ from 
calorimeter towers in a cone of $\Delta R = \sqrt{\Delta\eta^2 + 
\Delta\phi^2}\leq$~0.4 around the lepton is less than 10\% of the 
electron $\Et$ or muon $\pt$. In order to preserve pairs of leptons 
in close proximity to one another, if an additional muon or electron 
candidate is found within the $\Delta R \leq$~0.4 cone, calorimeter 
towers associated with it are not included in the $\Et$ sum.
An explicit requirement that the $\Delta R$ among all the reconstructed
leptons is greater than 0.05 guarantees that any two different leptons
are not based on the same track.

Electron candidates are required to have a ratio of HAD-to-EM energy 
consistent with an electromagnetic shower and are referred to as 
either \emph{central} or \emph{forward}, depending on whether they 
are identified within the central or forward calorimeter. Central 
electron identification requires a high-quality charged particle in 
the COT with $\pt \geq$~10~GeV/$c$, projecting to the geometrical 
acceptance of the central SMX detector, and matched to an EM energy 
cluster in the central calorimeter. Central electron candidates are 
selected using a likelihood method to combine electron identification 
variables into a single discriminant. A forward electron candidate 
is required to be detected within the geometrical acceptance of 
the forward SMX detector and to be associated with energy deposits 
consistent with those expected for an electron in both the forward 
calorimeter towers and SMX detector.  In order to reduce background 
from photons matched to misreconstructed calorimeter-seeded tracks, 
for each forward electron candidate we also require that the matching 
calorimeter-seeded track is consistent with a track formed only from 
hits in the silicon detector. A forward electron candidate that 
fails one or more of these requirements can still be selected using 
a likelihood-based method similar to that used for central electron 
selection.

Forward ($\eta_{\mathrm{det}}\geq$~1.2) muon reconstruction 
incorporates strict requirements on the number of COT hits and the 
$\chi^2$ of the track fit to suppress background from in-flight 
decays of pions and kaons.  The track's point of closest approach 
is also required to be consistent with the $\ppbar$ interaction 
point to suppress background from cosmic rays.

We also identify charged lepton candidates from reconstructed tracks, 
which neither geometrically extrapolate to the instrumented region 
of the calorimeter nor match to track stubs in the muon detectors.  
Such track-based candidates are required to satisfy the same quality 
requirements applied to the stubless muon candidates in the region 
$|\eta_{\mathrm{det}}|\leq$~1.2. Due to the lack of calorimeter 
information, electrons and muons cannot be reliably differentiated 
in this region, and these lepton candidates are therefore treated 
as being of either lepton flavor in the $Z$ candidate selection.  
Electron or track-lepton candidates are rejected if they are 
consistent with having originated from a photon conversion, as 
indicated by the presence of an additional nearby track.

The efficiencies for the aforementioned lepton selection criteria 
are evaluated in data and Monte Carlo simulation using inclusive 
$Z\to\ell\ell$ event samples.  The ratio of the efficiencies 
determined from the simulated and collision data samples is applied 
as a correction factor to the modeled rates of the contributing 
background and \ZZ~signal processes.

To identify the presence of neutrinos we define the missing 
transverse energy as $\MET~=~|\sum_i E_{T,i} \hat{n}_{T,i}|$, where 
$\hat{n}_{T,i}$ is the transverse component of the unit vector 
pointing from the interaction point to calorimeter tower $i$. The 
$\MET$ is corrected for the momentum of muons, which do not deposit 
all of their energy in the calorimeters, and for tracks that 
extrapolate to uninstrumented regions of the calorimeters.

Collimated clusters of particles ({\it jets}) are reconstructed from 
energy deposits in the calorimeters using the \textsc{jetclu} cone 
algorithm~\cite{jetclu} with a clustering radius of $\Delta R \equiv 
\sqrt{\Delta\eta^2 + \Delta\phi^2}=$~0.4. Their measured energies are 
corrected to match, on average, that of the showering parton using 
standard techniques~\cite{jet_corr}. Jets are selected if they have 
$\Et \geq$~15~GeV and $|\eta|\leq$~2.4.

\section{\ZZ~$\to$~$\llll$ analysis \label{sec:4leps}}

\subsection{Event selection \label{ssec:4l_evsel}}

The \ZZ~$\to$~$\llll$ candidate events are required to have exactly four 
reconstructed leptons with $\pt \geq$~10~GeV/$c$, at least one of which 
must have $\pt \geq$~20~GeV/$c$ and satisfy the trigger requirements. 
The leptons are grouped into opposite-charge and same-flavor pairs, 
treating the track-only lepton candidates as either electrons or muons, 
with the objective of identifying the leptonic decay products from 
each $Z$ boson decay. If an event has more than one possible pairing 
combination, the one for which the invariant masses of the two 
dilepton pairs lie closest to the known $Z$ boson mass~\cite{pdg}, 
$M_Z$, is chosen by minimizing $f(M_{1,2},M_{3,4}) = (M_{1,2}-M_Z)^2 + 
(M_{3,4}-M_Z)^2$, where $M_{1,2}$, $M_{3,4}$ are the two reconstructed 
dilepton masses.  One pair of leptons is required to have a reconstructed 
invariant mass within 15~GeV/$c^2$ of the known $Z$ boson mass, while 
the other is required to be within 50~GeV/$c^2$. The \ZZ~$\to$~$\llll$ 
acceptance is determined from a \textsc{pythia}-based Monte Carlo 
simulation~\cite{Sjostrand:2006za} followed by a \textsc{geant}-based 
simulation of the CDF II detector~\cite{geant}. The \textsc{cteq5l} 
parton distribution functions (PDFs) are used to model the momentum 
distribution of the initial-state partons~\cite{Lai:1999wy}.

\subsection{Background estimation \label{ssec:4l_background}}

The only significant background contribution to the $\llll$ final state is 
Drell-Yan (DY) production of a single $Z/\gamma^*$ boson in association 
with additional parton jets or photons that are misidentified as two 
additional leptons in the detector ({\it fakes}).  A data control sample 
is relied upon for estimating this contribution, since the simulation 
is not expected to accurately model the detector effects leading to 
the misidentification of showering partons as leptons. In event samples 
collected with jet-based triggers, we measure the probability for a jet 
to be identified as a lepton, correcting for the contribution of prompt 
leptons originating from $W$ and $Z$ boson decays.  The misidentification 
rate is measured as a function of lepton transverse energy, pseudorapidity, 
and flavor \cite{thesis}.  Data events with three identified leptons and 
a lepton-like jet \cite{fakes_foot}, 3$\ell$+$j_\ell$, and 
two identified leptons and two lepton-like jets, 2$\ell$+$2j_\ell$ that 
satisfy all other selection criteria are weighted by the measured 
misidentification rates associated with each lepton-like jet to provide 
an estimate of the background contribution. A $\mathcal{O}(1\%)$ correction 
is applied to account for double-counting due to the fraction of observed 
$3\ell$+$j_\ell$ events that originate from $2\ell$+2$j_\ell$ events where 
a single lepton-like jet is identified as a lepton.

\begin{figure*}[!hp]
\centering
{\includegraphics[width=.30\textwidth]{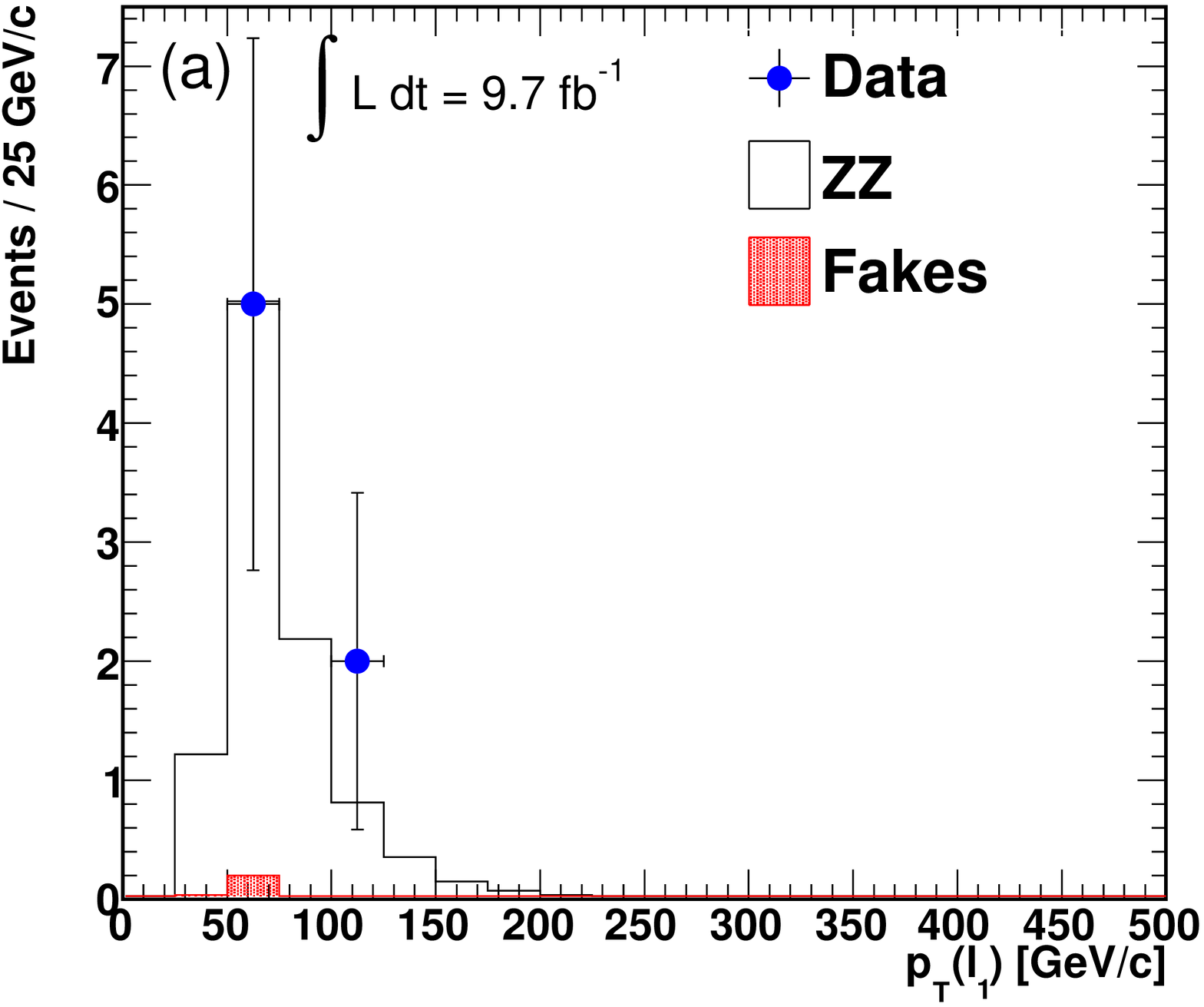}}
{\includegraphics[width=.30\textwidth]{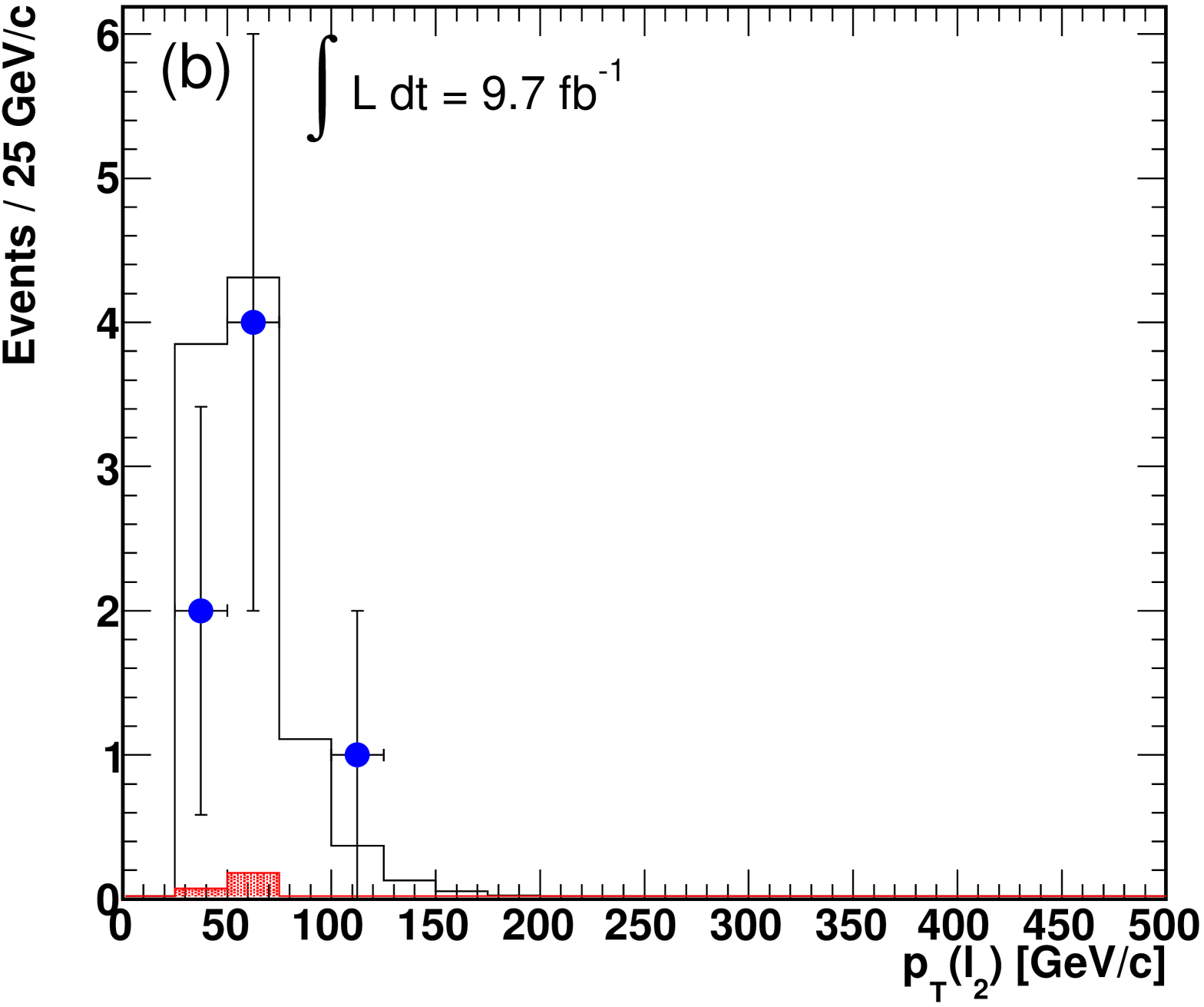}}
{\includegraphics[width=.30\textwidth]{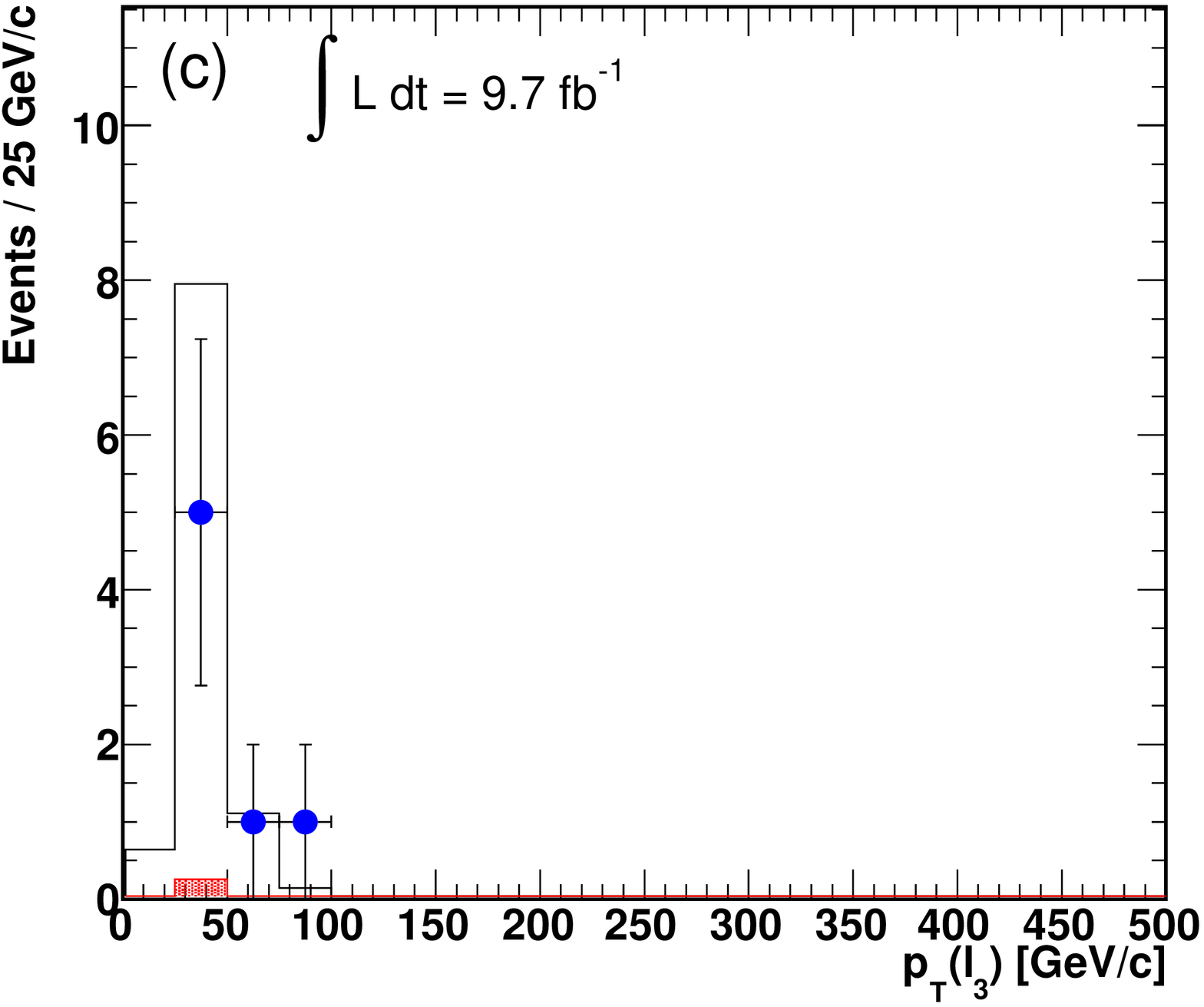}}\\
{\includegraphics[width=.30\textwidth]{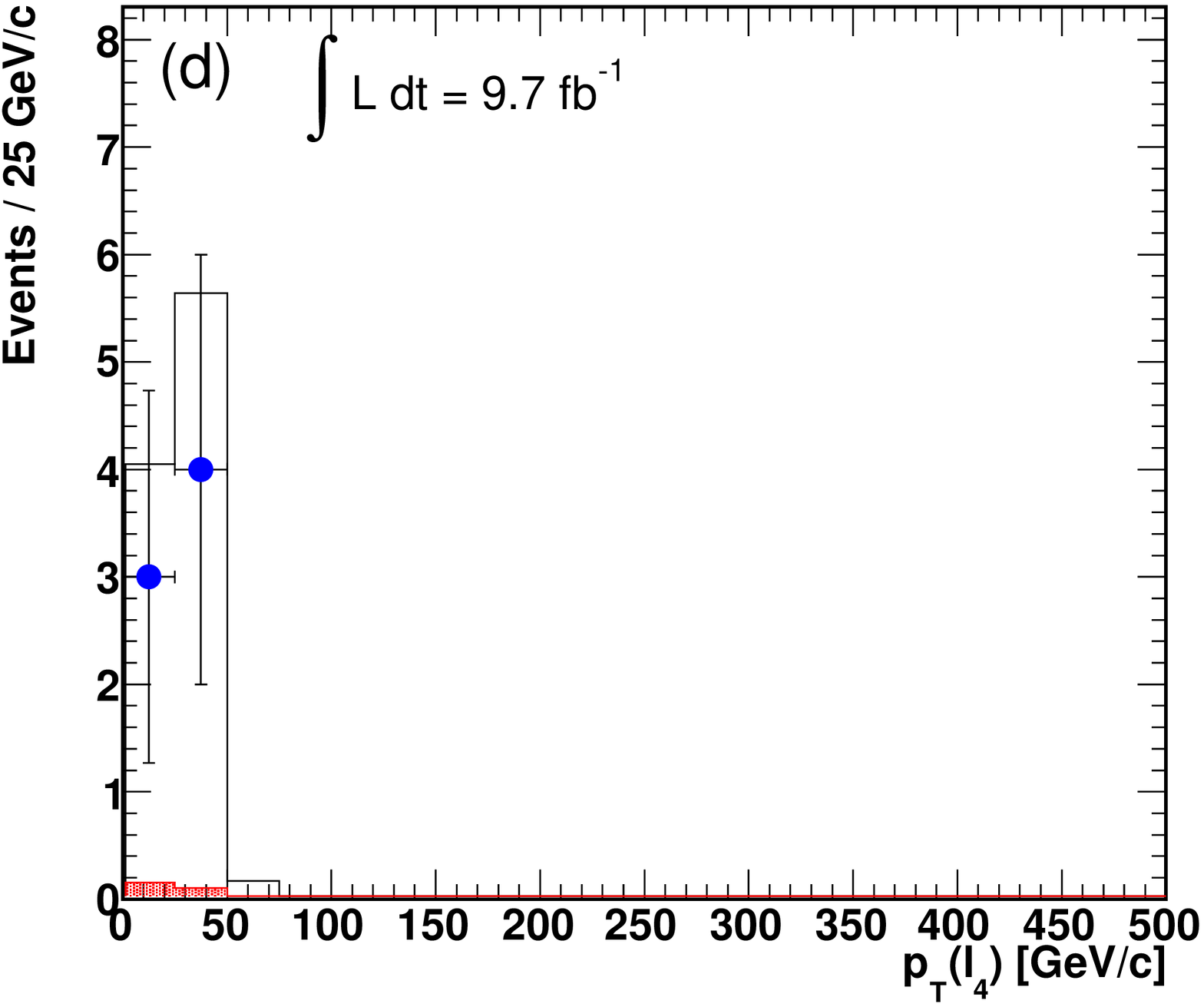}}
{\includegraphics[width=.30\textwidth]{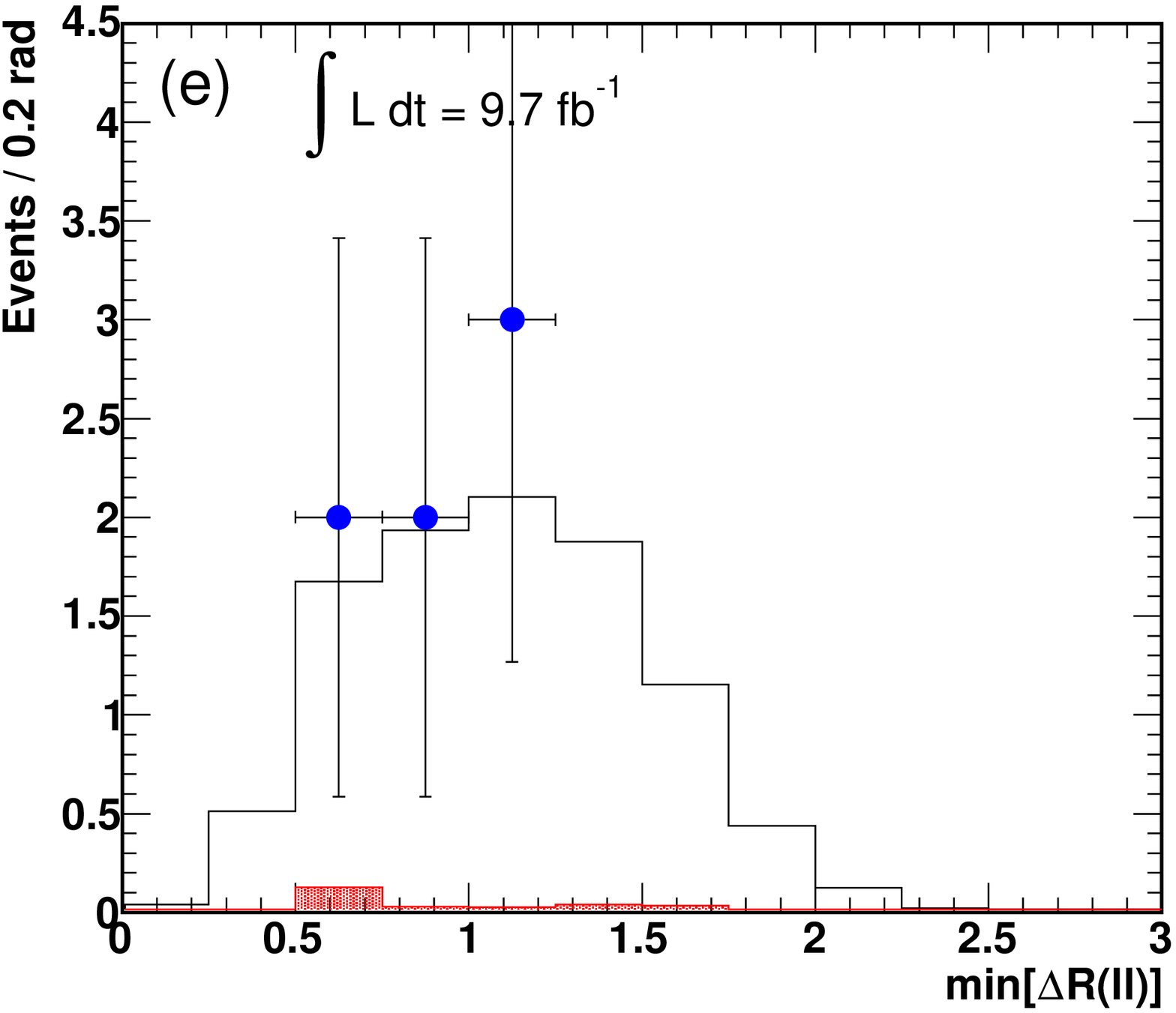}}
{\includegraphics[width=.30\textwidth]{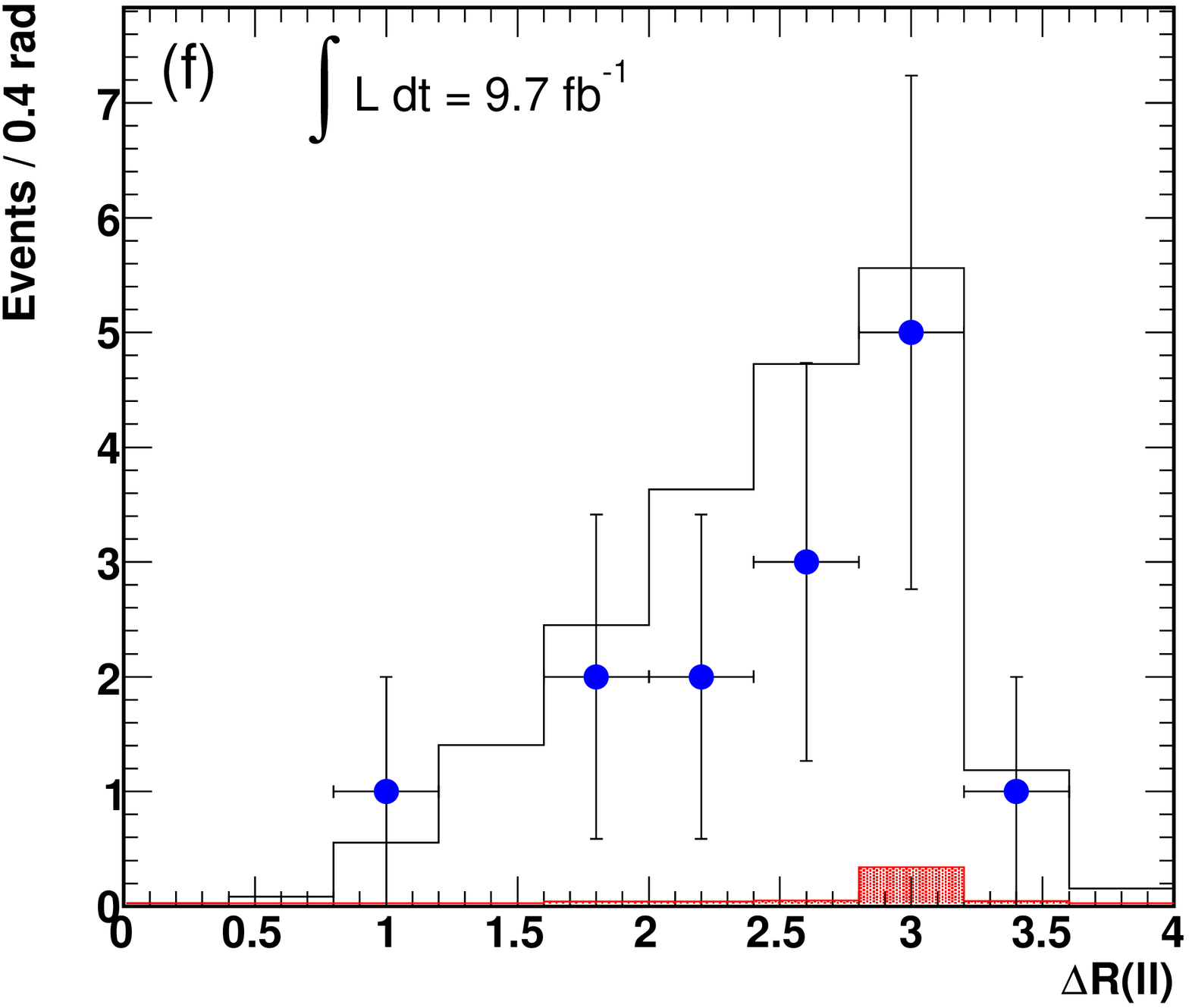}}\\
{\includegraphics[width=.30\textwidth]{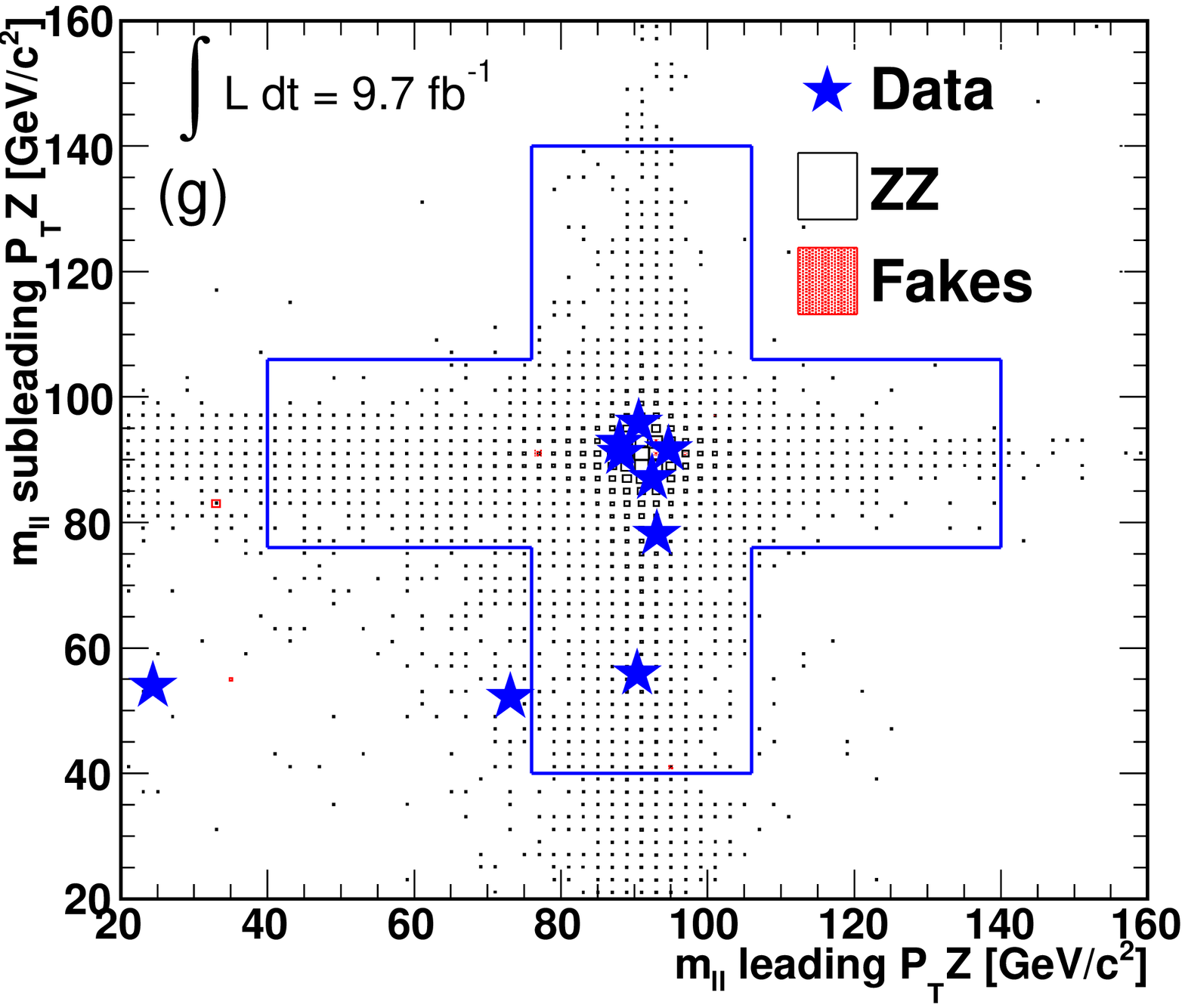}}
{\includegraphics[width=.30\textwidth]{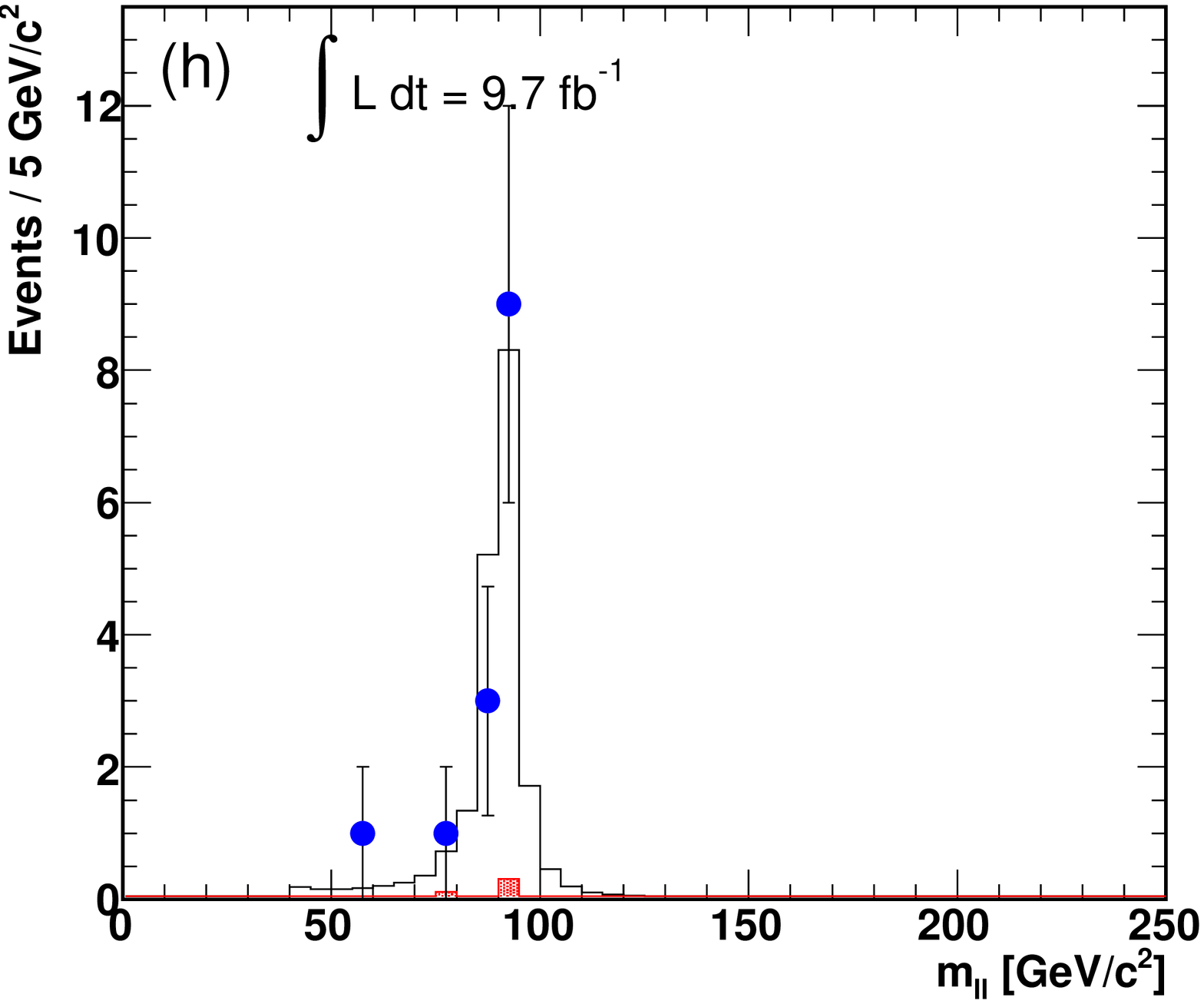}}
{\includegraphics[width=.30\textwidth]{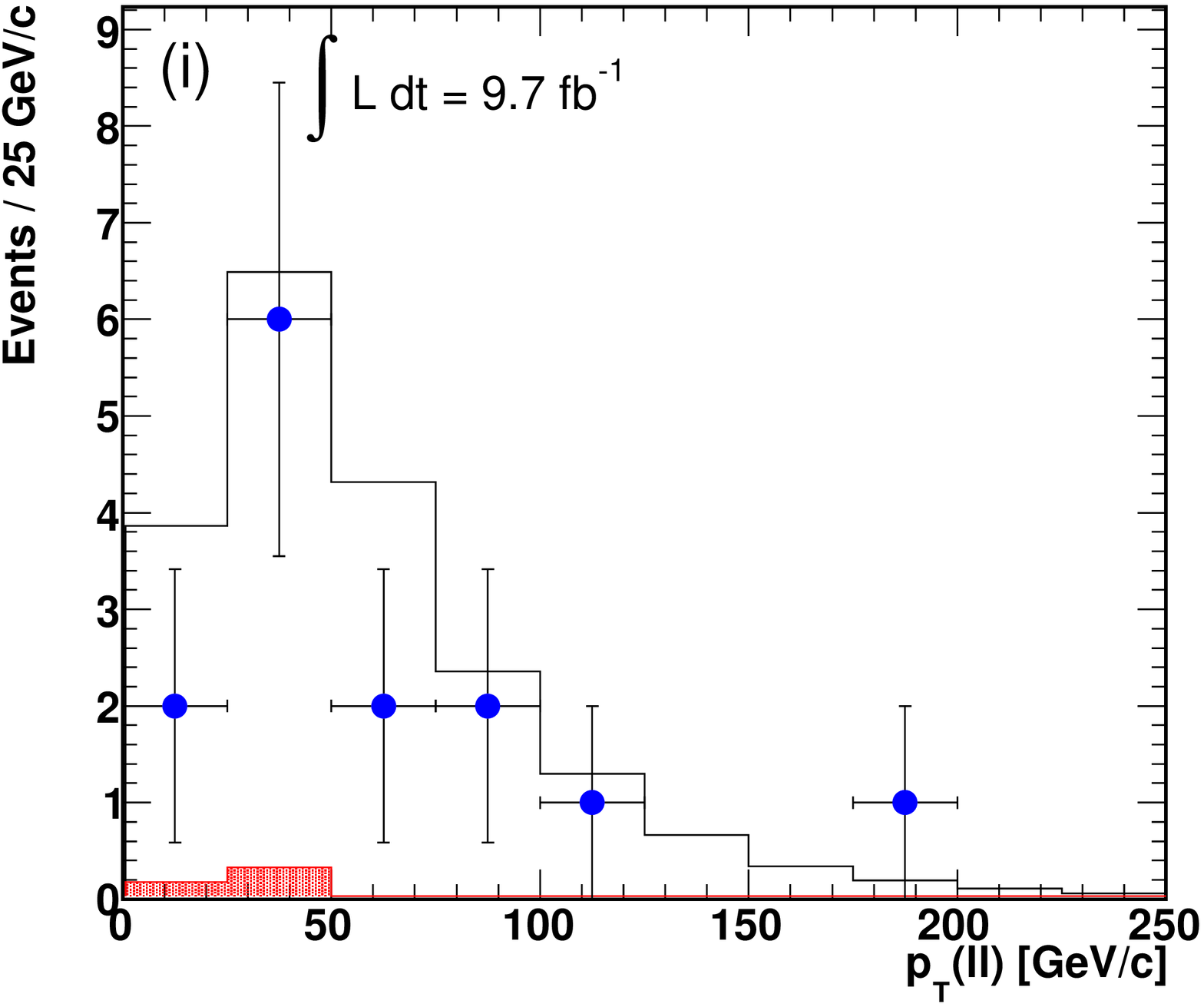}}\\
{\includegraphics[width=.30\textwidth]{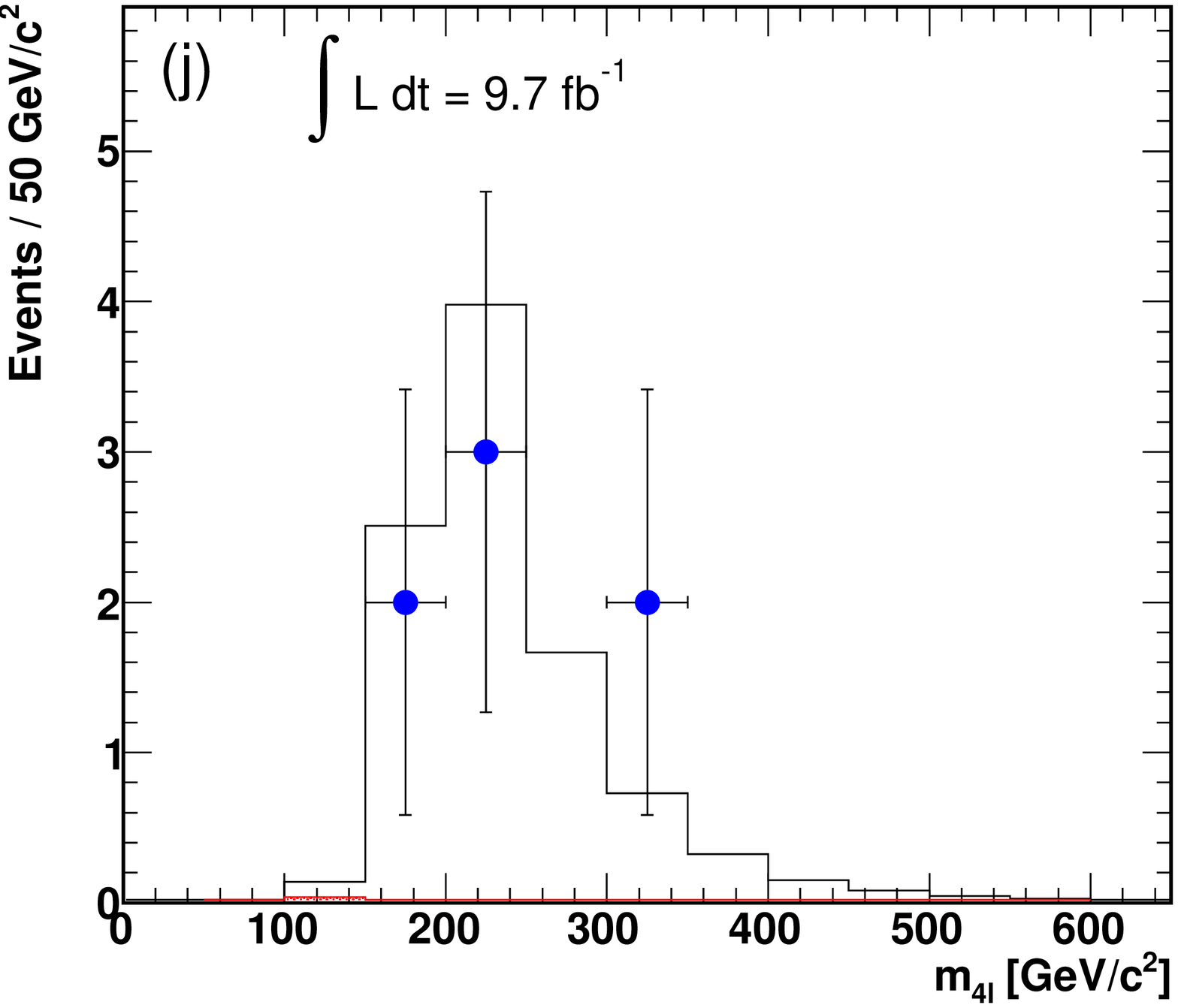}}
{\includegraphics[width=.30\textwidth]{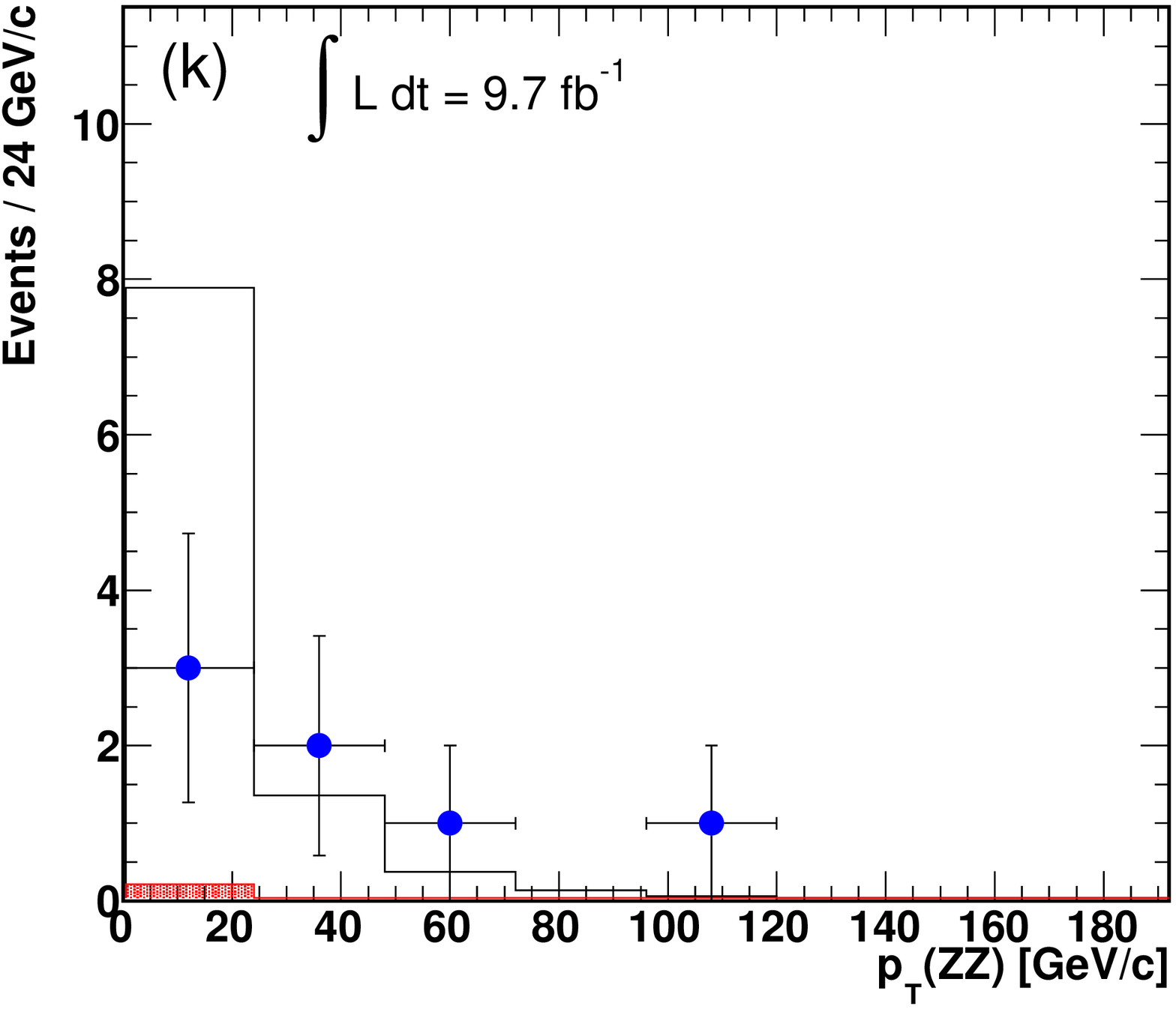}}
{\includegraphics[width=.30\textwidth]{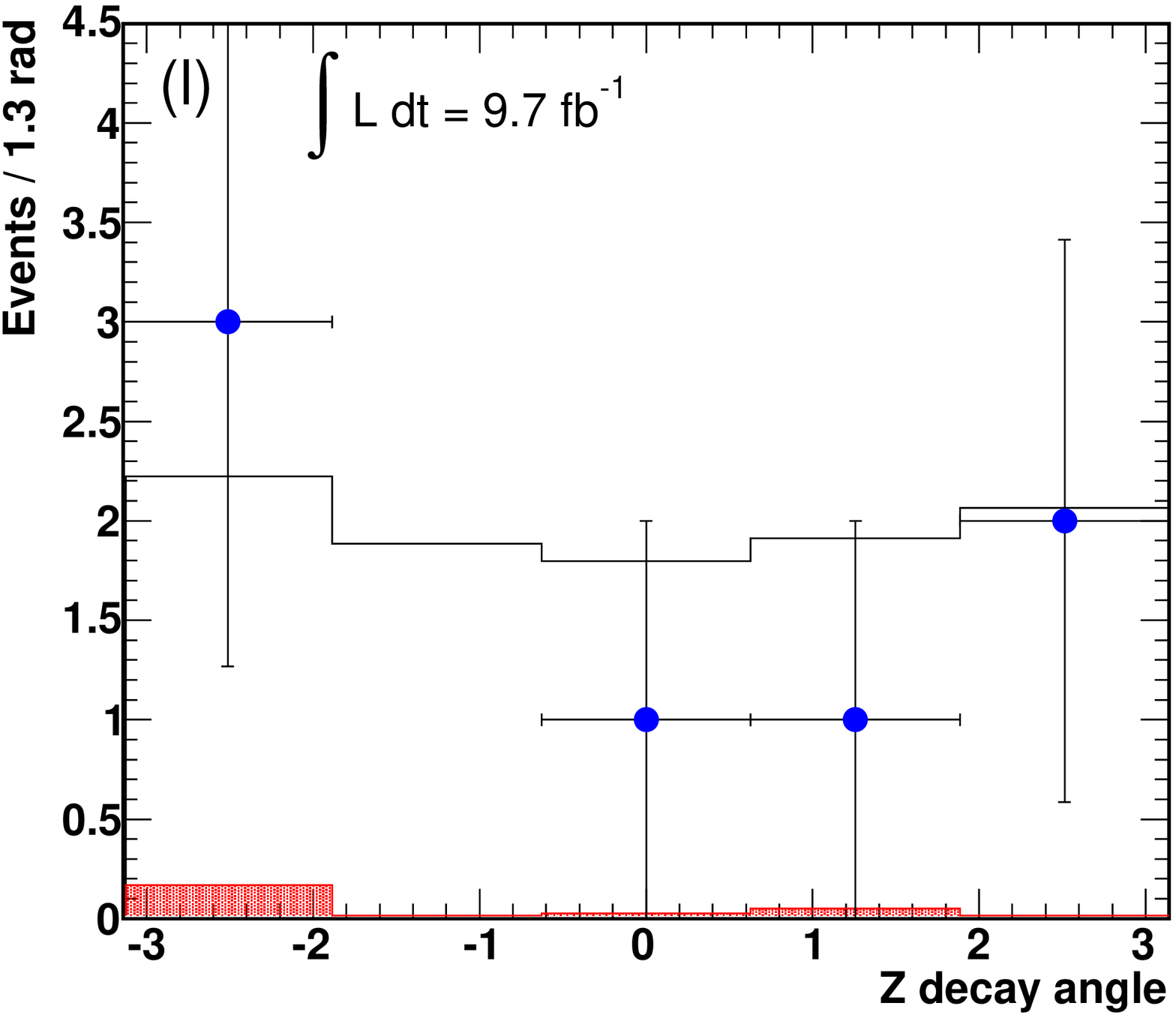}}\\
\caption{Comparisons of predicted and observed distributions of 
kinematic variables in events passing the full $\llll$ selection 
criteria: (a) transverse momentum of the leading lepton, 
(b) transverse momentum of the subleading lepton, (c) transverse 
momentum of the subsubleading lepton, (d) transverse momentum of 
the subsubsubleading lepton, (e) minimum $\Delta R$ between all 
possible lepton pairings, (f) opening angle $\Delta R$ between 
the two $Z$ boson candidates, (g) scatter distribution of the 
two reconstructed dilepton invariant masses, (h) invariant masses 
of lepton pairs associated with both $Z$ boson candidates, 
(i) transverse momenta of both $Z$ boson candidates, (j) four 
lepton invariant mass, (k) transverse momentum of the \ZZ~system, 
and (l) opening angle between the two $Z$ boson candidate decay 
product planes.}
\label{fig:ZZ4l_kin}
\end{figure*}

Table~\ref{tab:4l_exp} summarizes expected and observed event yields 
for the full data sample, corresponding to 9.7~$\fb$ of integrated 
luminosity.  Comparisons of the predicted and observed distributions 
of the most relevant kinematic variables in events passing the full 
$\llll$ selection criteria are shown in Fig.~\ref{fig:ZZ4l_kin}.  The 
agreement between the predicted and observed distributions indicates 
that the observed events are compatible with having originated from 
\ZZ~production.

\begin{table}[!htb]
\begin{center}
\caption{Predicted and observed numbers of \ZZ~$\to$~$\llll$ candidate 
events for the full CDF~II data sample.  The uncertainties on the 
predictions include both statistical and systematic contributions 
added in quadrature.}
\label{tab:4l_exp}
\begin{ruledtabular}
\begin{tabular}{lc}
Process         & Yield \\
\hline
\ZZ             & 9.59 $\pm$ 1.55\\
DY              & 0.06 $\pm$ 0.03\\
\hline
Total expected  & 9.65 $\pm$ 1.55\\
\hline
Data            & 7 \\
\end{tabular}
\end{ruledtabular}
\end{center}
\end{table}

\subsection{Systematic uncertainties \label{ssec:4l_sys}}

We account for sources of systematic uncertainty on the simulated detector 
acceptance of the signal and modeling of the background processes.  We 
assign a 2.5\% uncertainty on the \ZZ~signal acceptance from higher-order 
amplitudes not included in the simulation, comparing the acceptance of 
a NLO calculation with the leading order (LO) simulation used for this 
analysis.  In addition, a 2.7\% uncertainty is assigned to cover variations 
among different PDF model inputs on the simulation.  We also assign a 5.9\% 
uncertainty to the measured integrated luminosity~\cite{Acosta:2002hx}, 
accounting for both the uncertainties on the acceptance and operation of 
the luminosity monitor and the measurement of the total $\ppbar$ cross 
section at the Tevatron.  The limited size of the sample used to derive 
lepton-identification efficiencies results in an additional 3.6\% 
uncertainty on simulated acceptances, obtained through the propagation 
of this uncertainty to the efficiencies on the total acceptance.  Due 
to the presence of four leptons in each event and the high efficiencies 
($\simge$~90\%) of the single lepton triggers, uncertainties on measured 
trigger efficiencies have a minimal ($\approx$~0.04\%) impact on the 
overall acceptance uncertainties.  The uncertainty on the DY background 
contribution is evaluated by applying a range of lepton misidentification 
rates as measured from the control samples obtained using different trigger 
requirements. The resulting uncertainty on the background contribution is 
50\%. This has a negligible effect on the precision of the cross section 
measurement due to the small size of the predicted contribution.

\subsection{Result \label{ssec:4l_result}}

To extract the \ZZ~production cross section, a Bayesian method is 
employed~\cite{stats}, building a likelihood function that takes 
as inputs the expected signal acceptance, the number of expected 
background events, and the number of observed events passing the 
selection criteria described above.  The resulting expression 
gives the Poisson probability for obtaining the observed number 
of events as a function of the \ZZ~production cross section, 
$\sigma(\ppbar \to \ZZ)$, to which we assign a uniform prior 
probability over the range of non-negative values.  The function 
also includes terms for truncated, Gaussian-constrained nuisance 
parameters corresponding to each systematic uncertainty source, 
which are integrated over their parameter space. The value 
of the cross section, relative to the SM expectation, that 
maximizes this probability is the result of the measurement, 
for which we obtain $\sigma(\ppbar\to \ZZ)/\sigma^{\rm SM-NLO} 
=$~0.73~$^{+0.31}_{-0.24}$~(stat)~$^{+0.08}_{-0.05}$~(syst), 
which corresponds to a value of $\sigma(\ppbar\to \ZZ) =
$~0.99~$^{+0.45}_{-0.35}$~(stat)~$^{+0.11}_{-0.07}$~(syst)~pb 
in the zero-width approximation.


\section{\ZZ~$\to$~$\llvv$ analysis \label{sec:llvv}}

The \ZZ~$\to$~$\llvv$ decay mode has a slightly larger branching ratio 
(approximately 3\%).  The two neutrinos produced in the decay of one 
$Z$ boson cannot be directly detected, and their presence is inferred 
from the presence of significant $\MET$.  The \ZZ~$\to$~$\llvv$ 
candidate events are required to contain exactly two oppositely-charged 
and same-flavor leptons. One of the two leptons has to match the 
requirements of a single lepton trigger and have $\pt \geq$~20~GeV/$c$, 
while the second lepton is required to have $\pt \geq$ 10~GeV. The 
invariant mass of the dilepton pair is required to be within 15~GeV/$c^2$ 
of the known $Z$ boson mass~\cite{pdg}.

In Tevatron collisions the dominant source of dilepton events is 
inclusive DY production, which has a cross section many orders 
of magnitude larger than that of the signal process investigated 
here.  The main feature that distinguishes events associated with 
the two processes is the presence of significant $\MET$ within 
signal events.  Other background contributions come from the 
leptonic decays of \WW~and \WZ~boson pairs. In the $WW\to\ell\nu
\ell^{^{(_{\null'})}}\nu$ decay, a pair of leptons can be produced 
in association with a significant amount of $\MET$ due to the 
presence of the two neutrinos, while the $WZ\to\ell\nu
\ell^{^{(_{\null'})}+}\ell^{^{(_{\null'})}-}$ decay can produce 
a similar signature when one of the three leptons lies outside 
the detector coverage and is therefore undetected. Additional, 
non-negligible background contributions originate from $W\gamma$ 
and $W$+jets production, where the photons or jets are misidentified 
as leptons, and from $\ttbar$ quark pair-production.

\subsection{Event selection \label{ssec:llvv_selection}}

In order to extract the \ZZ~$\to$~$\llvv$ signal from the background-dominated 
event sample, we exploit differences in the kinematic properties of signal 
and background events.  Since \ZZ~$\to$~$\llvv$ events typically contain 
little additional hadronic activity, we veto events that have a jet with 
$\Et \geq$~15~GeV recoiling against the reconstructed $Z$ boson candidate 
($\Delta\phi(j,Z)\geq\pi/2$).  Events originating from DY production 
typically contain $\MET$ generated from mismeasured energies of jets 
recoiling against the $Z$ boson. By vetoing events containing these types 
of jets, the DY background contribution is significantly reduced with a 
minimal ($<$~5\%) impact on signal acceptance.  This requirement also 
suppresses potential signal contributions from \ZZ~$\to$~$\ell\ell$jj 
decays, whose kinematic event observables are different from those of 
the targeted leptonic decays. Less than a 2\% fraction of events remaining 
after this requirement contain a jet with $E_T\geq$~15~GeV, but the DY 
process is still the dominant contributor of background events to this 
sample.

\begin{figure}[!h]
\centering
\includegraphics[width=\columnwidth]{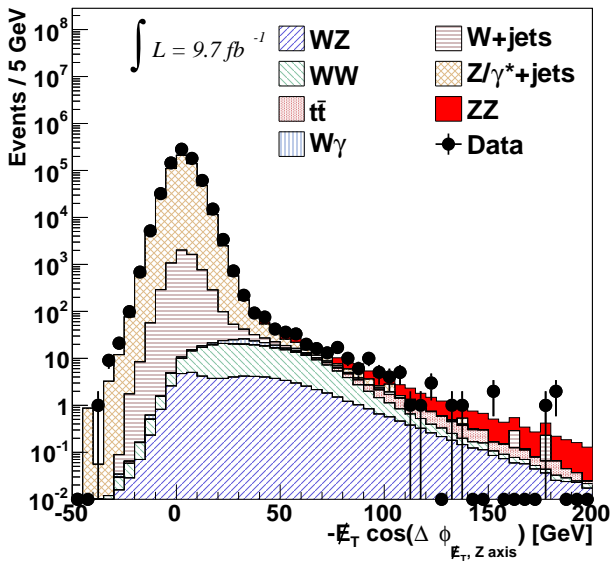}
\caption{Predicted and observed distributions of $\MET^{\rm ax}$ 
from selected events prior to the application of the $\MET^{\rm ax}$ 
requirement.  The observed data are overlaid on stacked predictions 
obtained from the modeling of contributing background and signal 
processes.}
\label{fig:MetAx}
\end{figure}

\begin{figure}[!h]
\centering
\includegraphics[width=\columnwidth]{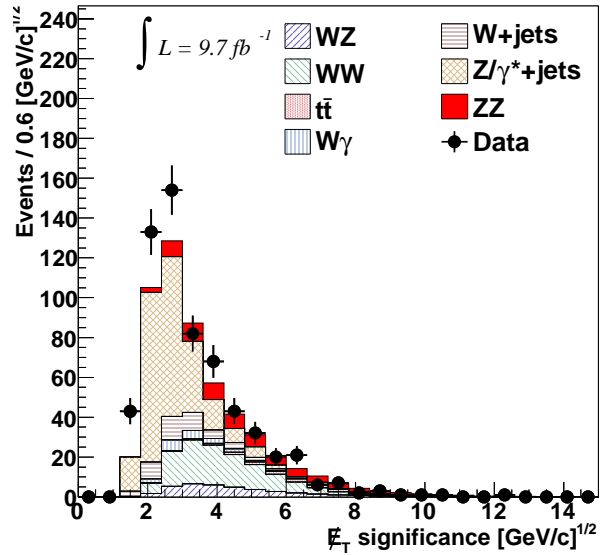}
\caption{Predicted and observed distributions of $\MET^{\rm sig}$ 
from selected events prior to the application of the $\MET^{\rm sig}$ 
requirement.  The observed data are overlaid on stacked predictions 
obtained from the modeling of contributing background and signal 
processes.}
\label{fig:MetSig_nocut}
\end{figure}

Further separation between signal and background contributions is 
achieved by requiring the $\MET$ to be antialigned with the direction 
of the reconstructed $Z$ boson $\pt$. We select events with 
$\MET^{\rm ax}\equiv - \MET \cos\Delta\phi(\hat{\MET},\hat{p}_T^Z)~\geq
$~30~GeV, where $\Delta\phi(\hat{\MET},\hat{p}_T^Z)$ is the angle between 
the \mbox{$\vec{E}\kern-0.50em\raise0.10ex\hbox{/}_{T}$} and the $Z$ 
boson $\pt$. The predicted and observed distributions of $\MET^{\rm ax}$ 
in selected events are shown in Fig.~\ref{fig:MetAx}.  The predicted 
distributions are those obtained using the modeling described in the 
following section.  This requirement rejects 99.8\% of the remaining 
DY background, while preserving approximately 30\% of the signal. To 
reduce the background contribution from processes not resulting in 
final-state neutrinos, in which $\MET$ is generated through detector 
mismeasurements, we also require the observed $\MET$ to be significant 
compared with the overall energy deposited in the calorimeter, 
$\MET^{\rm sig}\equiv \MET/\sqrt{\sum \Et} \geq$~3.0~GeV$^{1/2}$, 
where $\sum \Et$ represents the scalar sum of transverse energies 
deposited in the calorimeters. The predicted and observed 
distributions of $\MET^{\rm sig}$ in selected events are shown in 
Fig.~\ref{fig:MetSig_nocut}.

\begin{figure*}[!htbp]
\centering
{\includegraphics[width=0.30\textwidth]{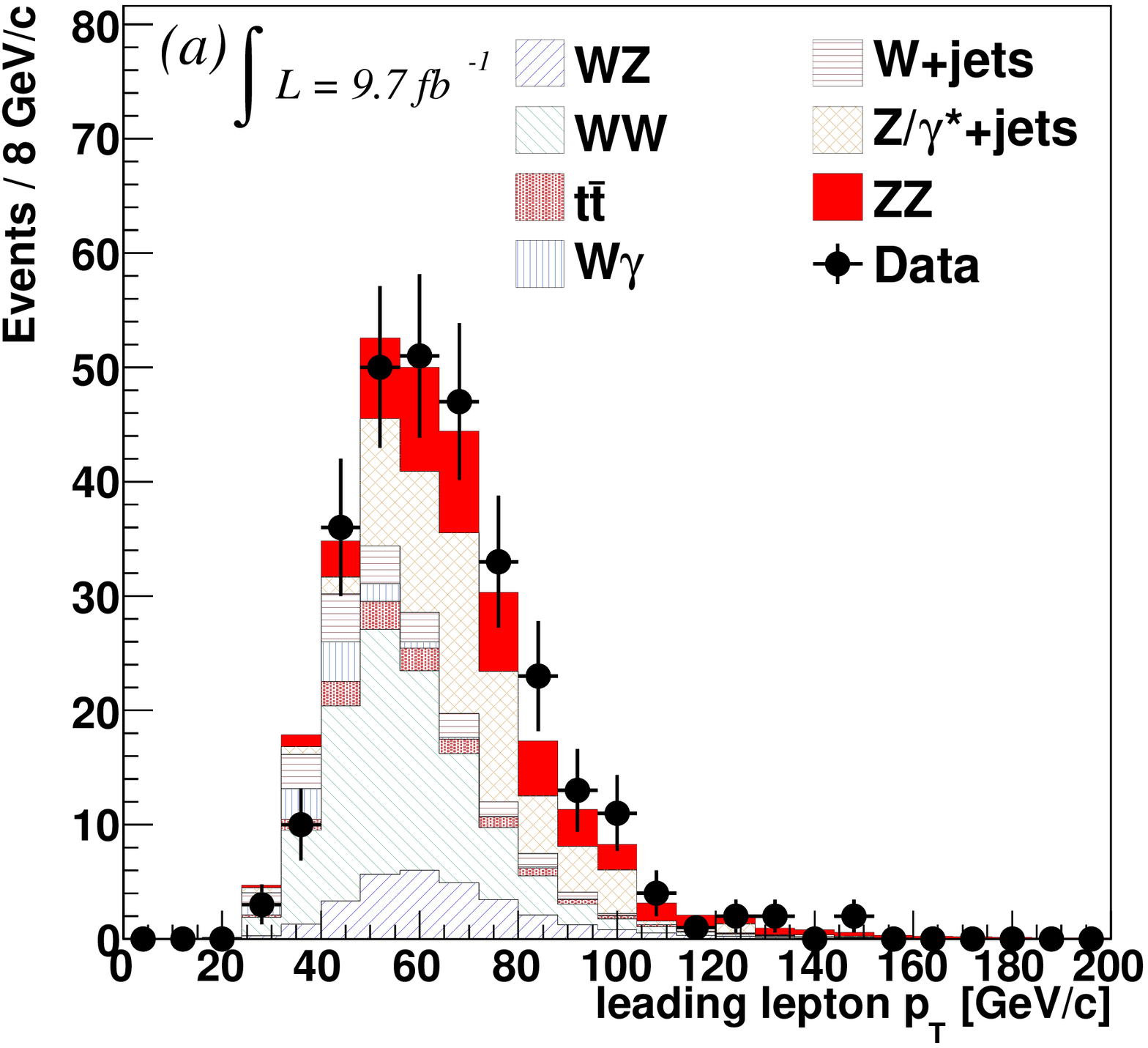}}
{\includegraphics[width=0.30\textwidth]{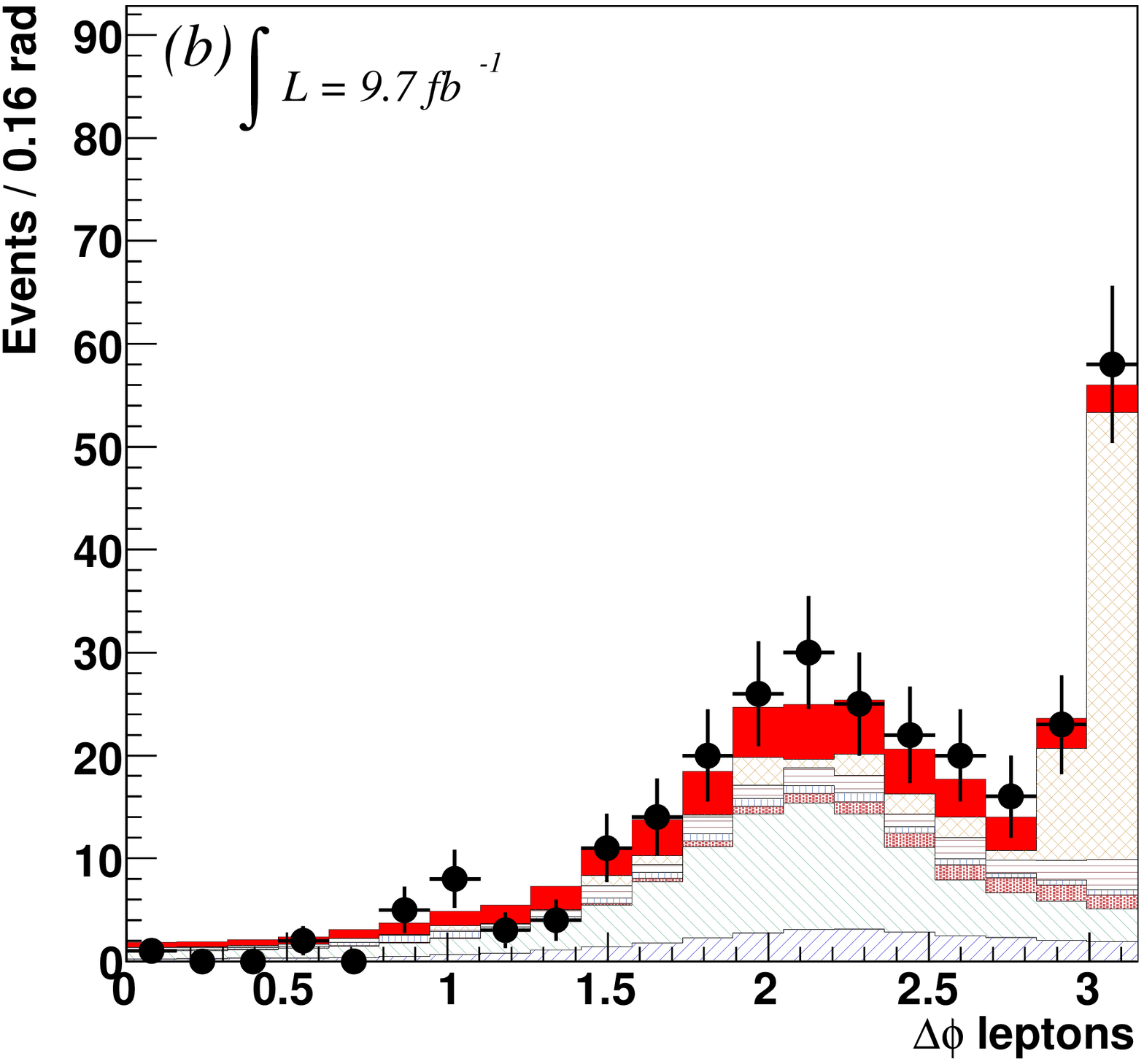}}
{\includegraphics[width=0.30\textwidth]{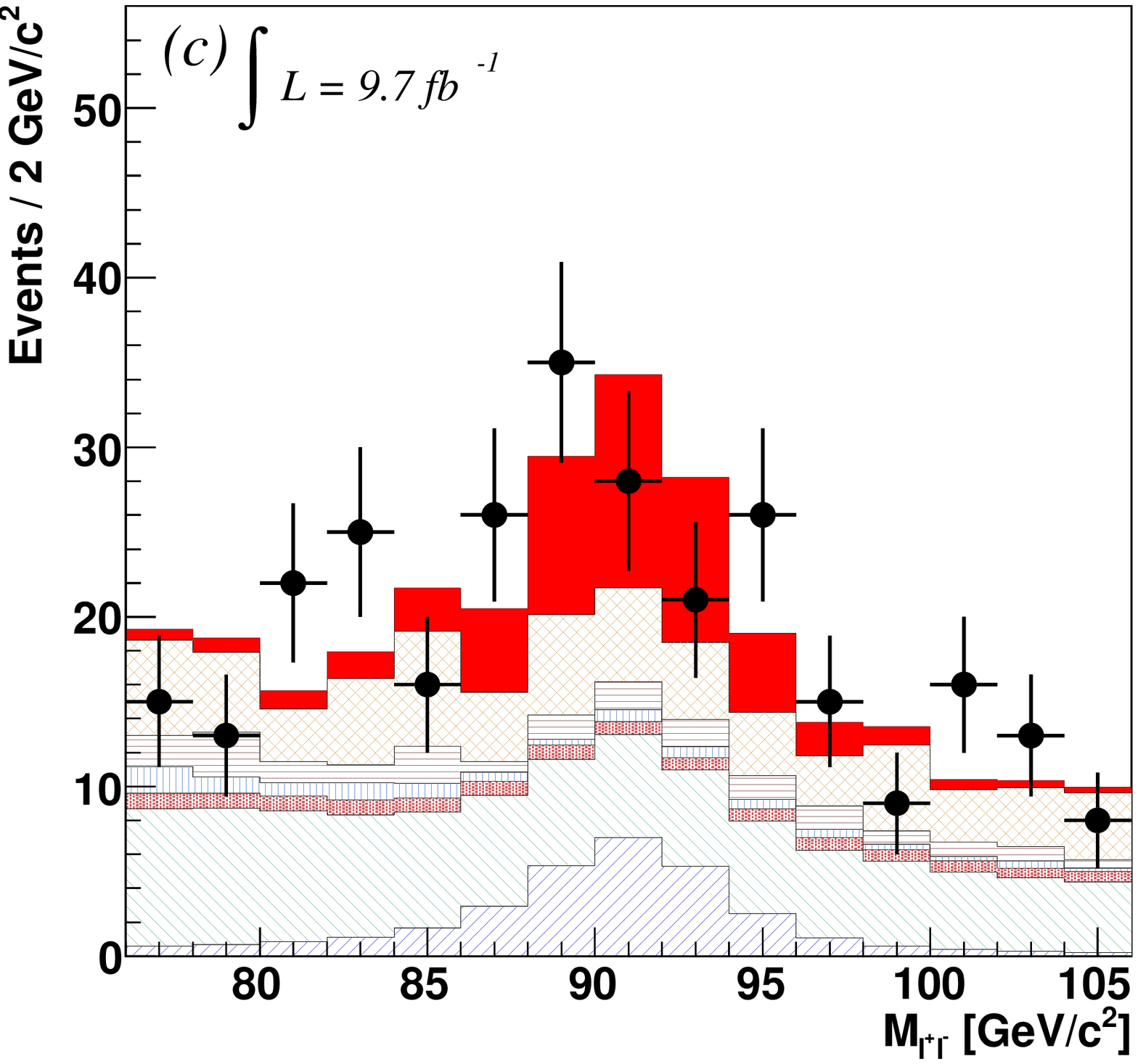}}
{\includegraphics[width=0.30\textwidth]{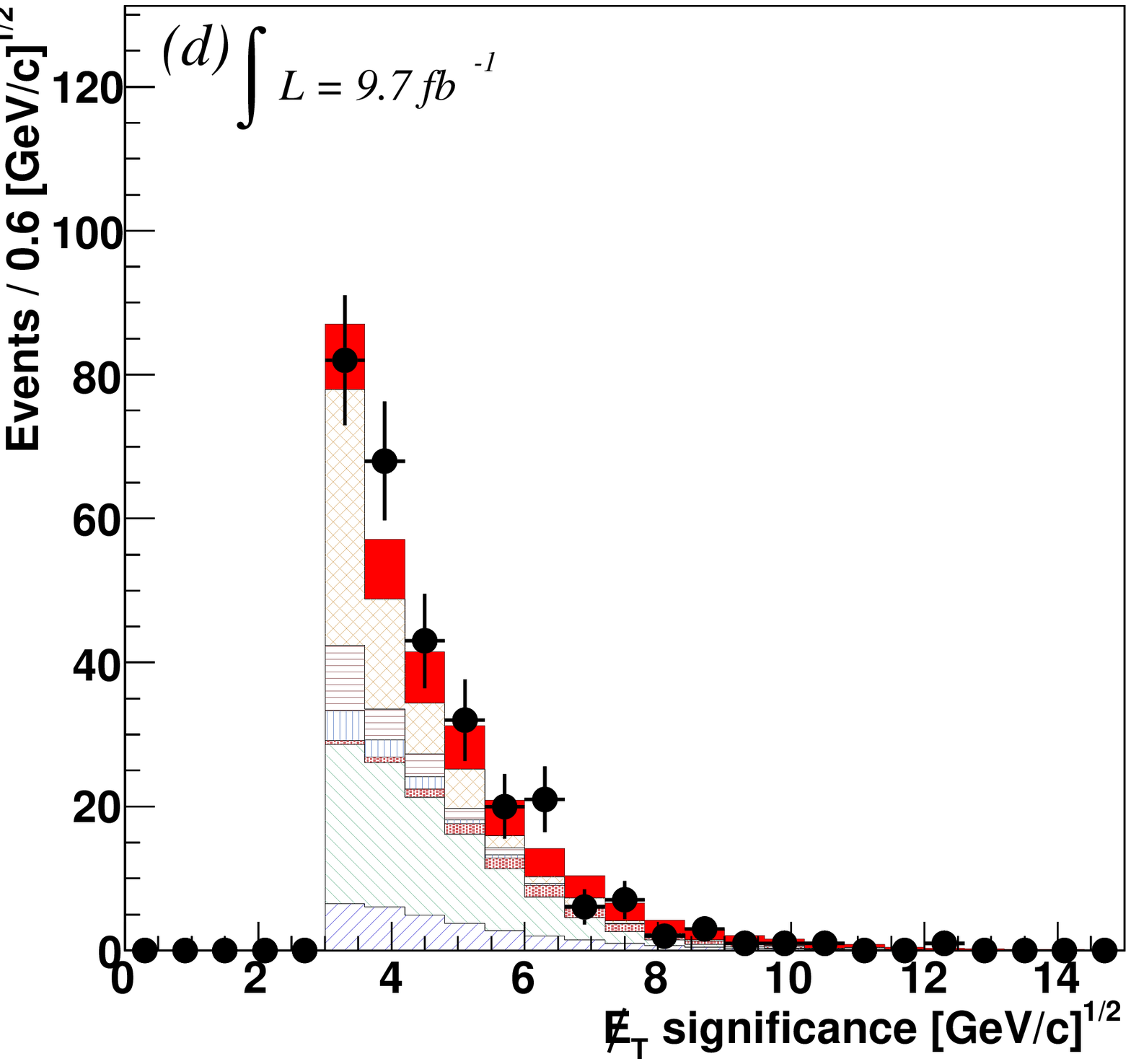}}
{\includegraphics[width=0.30\textwidth]{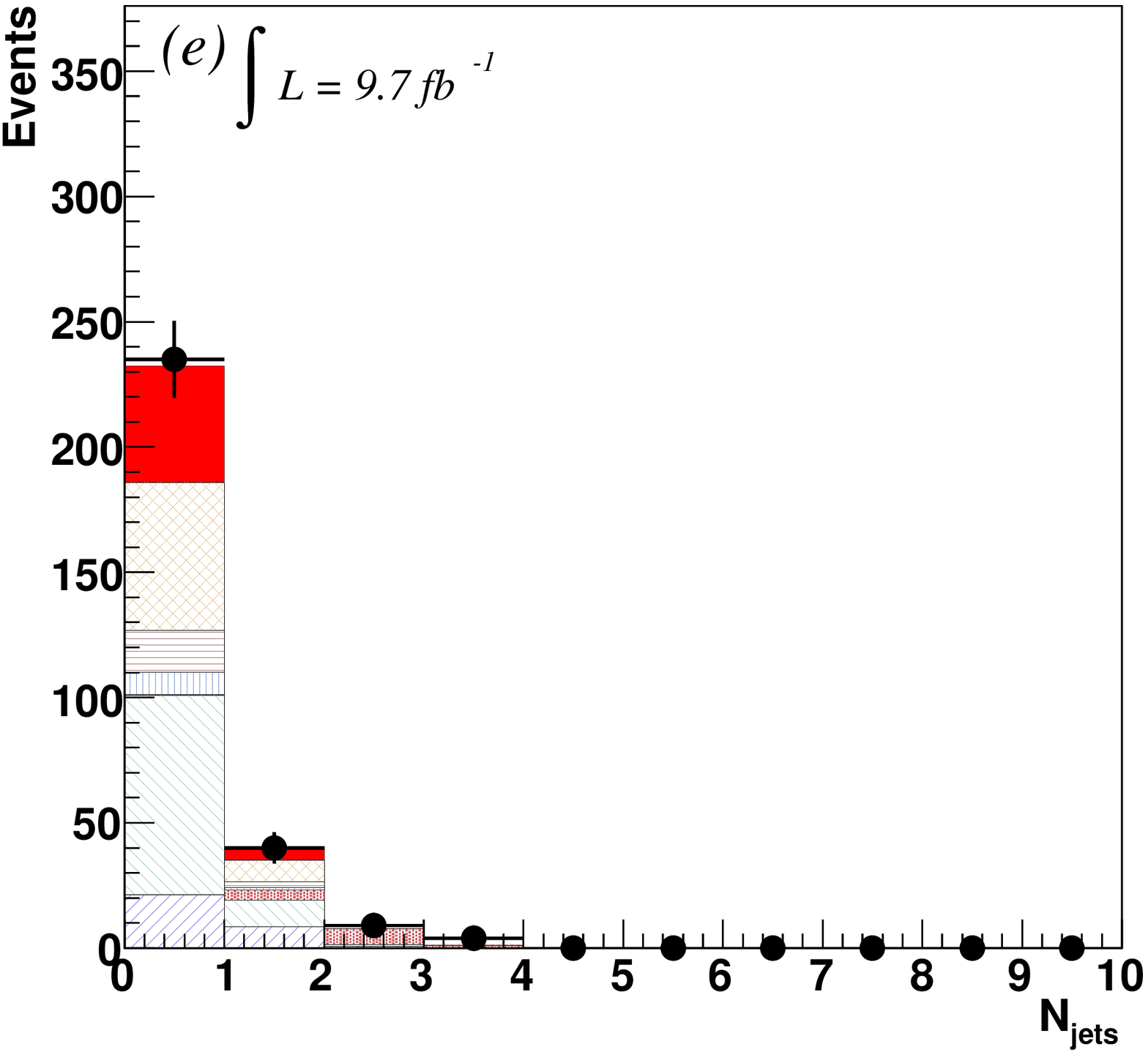}}
{\includegraphics[width=0.30\textwidth]{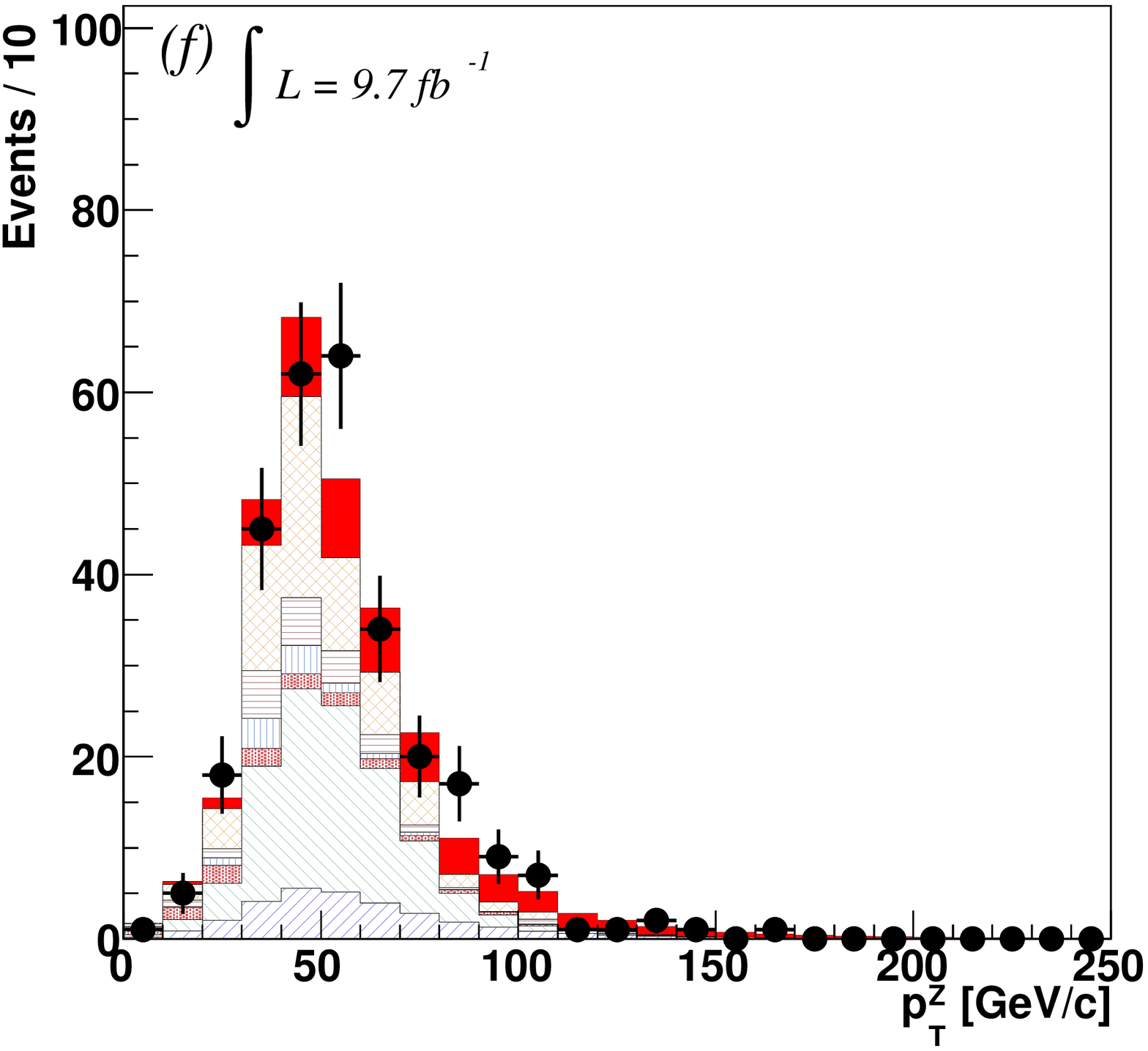}}
{\includegraphics[width=0.30\textwidth]{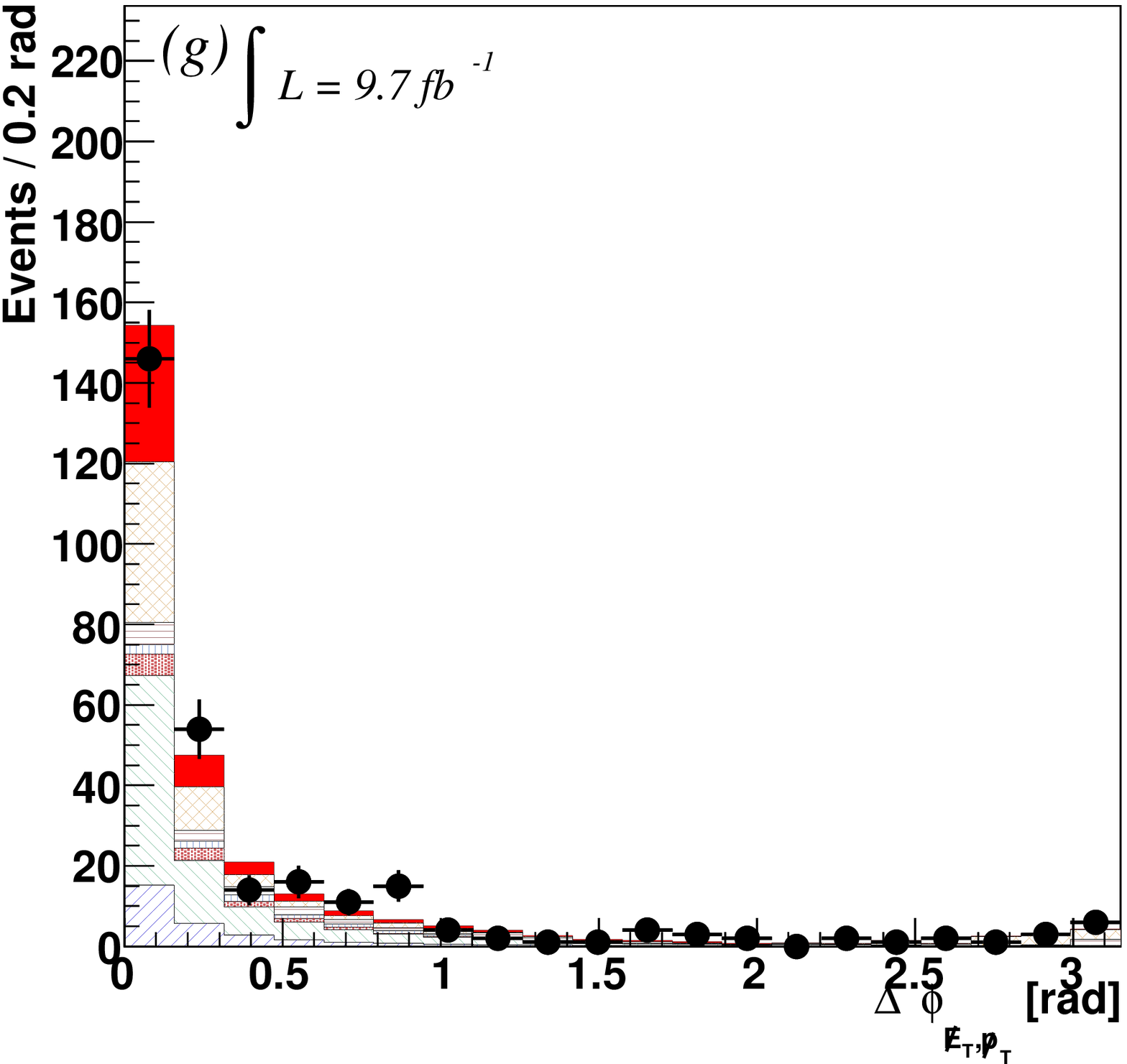}}
\caption{Comparisons of predicted and observed distributions of 
kinematic variables taken as inputs to the neural network for 
separating signal and background contributions in events passing 
the full $\llvv$ selection criteria: (a) transverse momentum of 
the leading lepton, (b) opening angle between the two leptons in 
the detector transverse plane, (c) reconstructed dilepton invariant 
mass, (d) $\MET$ significance, (e) number of reconstructed jets, 
(f) transverse momentum of the dilepton system, and (g) angle in 
the detector transverse plane between the $\MET$ and the $\MPT$.}
\label{fig:ZZllvv_kin}
\end{figure*}

\subsection{Background estimation \label{ssec:llvv_model}}

Modeling of signal and background processes contributing to 
the $\llvv$ final state is obtained primarily from simulation, 
similarly to what is done for the \ZZ~signal in the $\llll$ 
final state, using the \textsc{cteq5l}~\cite{Lai:1999wy} PDF 
model and a \textsc{geant}-based simulation of the CDF~II 
detector. The \ZZ, \WZ, DY, and $\ttbar$ processes are simulated 
using \textsc{pythia}, while the \WW~process is simulated using 
\textsc{mc@nlo}~\cite{Frixione:2002ik}.  The $W\gamma$ production 
is modeled with the Baur generator~\cite{Baur:1992cd}.  Simulated 
diboson and $\ttbar$ event samples are normalized to the 
highest-order theoretical cross section available~\cite{SMXS-old,
XSttbar}.  Normalization of the simulated DY sample is based on 
observed data in an independent control sample as described in 
more detail below.  The $W$+jets contribution is estimated using 
the same data-driven method used to estimate the DY background 
contribution to the $\llll$ final state. In this case, the same 
jet-to-lepton misidentification rates are applied as weights 
in events with one identified lepton and a lepton-like jet, 
1$\ell$+$j_\ell$.

Table~\ref{tab:ZZllvv_exp} summarizes predicted signal and 
background contributions to the sample after application of the 
selection criteria.  The background contributions from DY, \WW, 
and \WZ~production are reduced to levels comparable with that 
of the expected \ZZ~signal contribution. 

\begin{table}[!htb]
\begin{center}
\caption{Predicted and observed numbers of \ZZ~$\to$~$\llvv$ candidate 
events for the full CDF~II data sample. The uncertainties on the 
predictions include both statistical and systematic contributions 
added in quadrature.}
\label{tab:ZZllvv_exp}
\begin{ruledtabular}
\begin{tabular}{lc}
Process          & Yield \\
\hline
DY               & 67.2 $\pm$ 10.8 \\
$t\bar t$        & 11.5 $\pm$ 2.1 \\
$W+$jets         & 20.0 $\pm$ 5.3  \\
$W\gamma$        & 9.7  $\pm$ 1.2 \\
$WW$             & 91.2 $\pm$ 8.5  \\
$WZ$             & 30.4 $\pm$ 4.3 \\
\hline
Total background & 230 $\pm$ 15.5 \\
\hline
\ZZ              & 52.5 $\pm$ 9.2 \\
\hline
Data             & 288 \\
\end{tabular}
\end{ruledtabular}
\end{center}
\end{table}


The modeling of DY and \WW~background contributions is tested in 
independent data samples with kinematic properties similar to 
those of the signal sample.  DY background modeling is tested 
using a sample of events with $\MET^{\rm ax}\leq$ 25 GeV that 
otherwise satisfy the selection criteria of the signal sample 
with the exception of the requirement on $\MET^{\rm sig}$, 
which is not applied (low $\MET$ control sample).  Modeling 
of the \WW~background contribution is tested using a sample 
of $e^\pm\mu^\mp$ events passing the same requirements applied 
for signal events with the exceptions of no requirement on 
$\MET^{\rm sig}$ and restricting the dilepton invariant mass 
to the region 40~$\leq M_{e\mu}\leq$~140~GeV/$c^2$ ($e$--$\mu$ 
control sample).  The contribution of the \ZZ~signal process to 
each of these control samples is negligible, while the $e$--$\mu$
control sample contains a small, residual contribution from DY 
production through $Z\to\tau\tau$ decays.  A normalization for 
the predicted DY contribution to the signal sample is obtained 
from the $e$--$\mu$ control sample by fitting the prediction to 
data in the high $\MET$ region \cite{thesis}. In this kinematic 
region, the $\Delta\phi(\ell\ell)$ distributions of the \WW~and 
DY event contributions have different behaviors: the DY 
contribution, mainly coming from $Z\to\tau^+\tau^-$ decays, is 
peaked at $\Delta\phi(e\mu)\approx\pi$, while the contribution 
from \WW~production has a broader $\Delta\phi(e\mu)$ distribution 
peaked at $\approx$~2.  Hence, contributions from the two 
processes can be distinguished in this kinematic region, 
allowing for the extraction of a correction factor that can 
be applied to the predicted DY contribution in the signal 
sample.  

\subsection{Neural network separation \label{ssec:llvv_neuralnet}}

\begin{figure}[!htbp]
\centering
\subfigure[]{\includegraphics[width=\columnwidth]{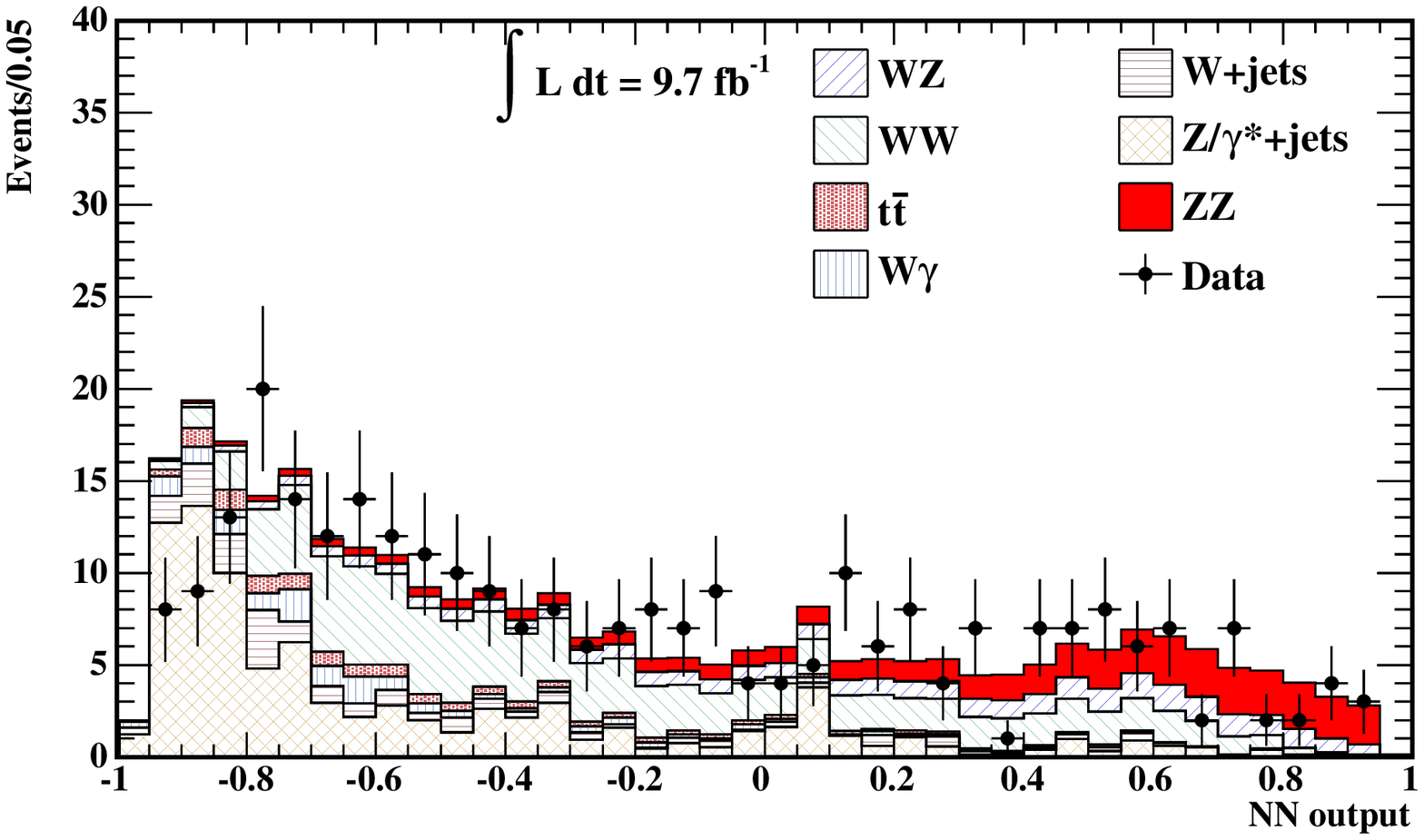}}
\subfigure[]{\includegraphics[width=\columnwidth]{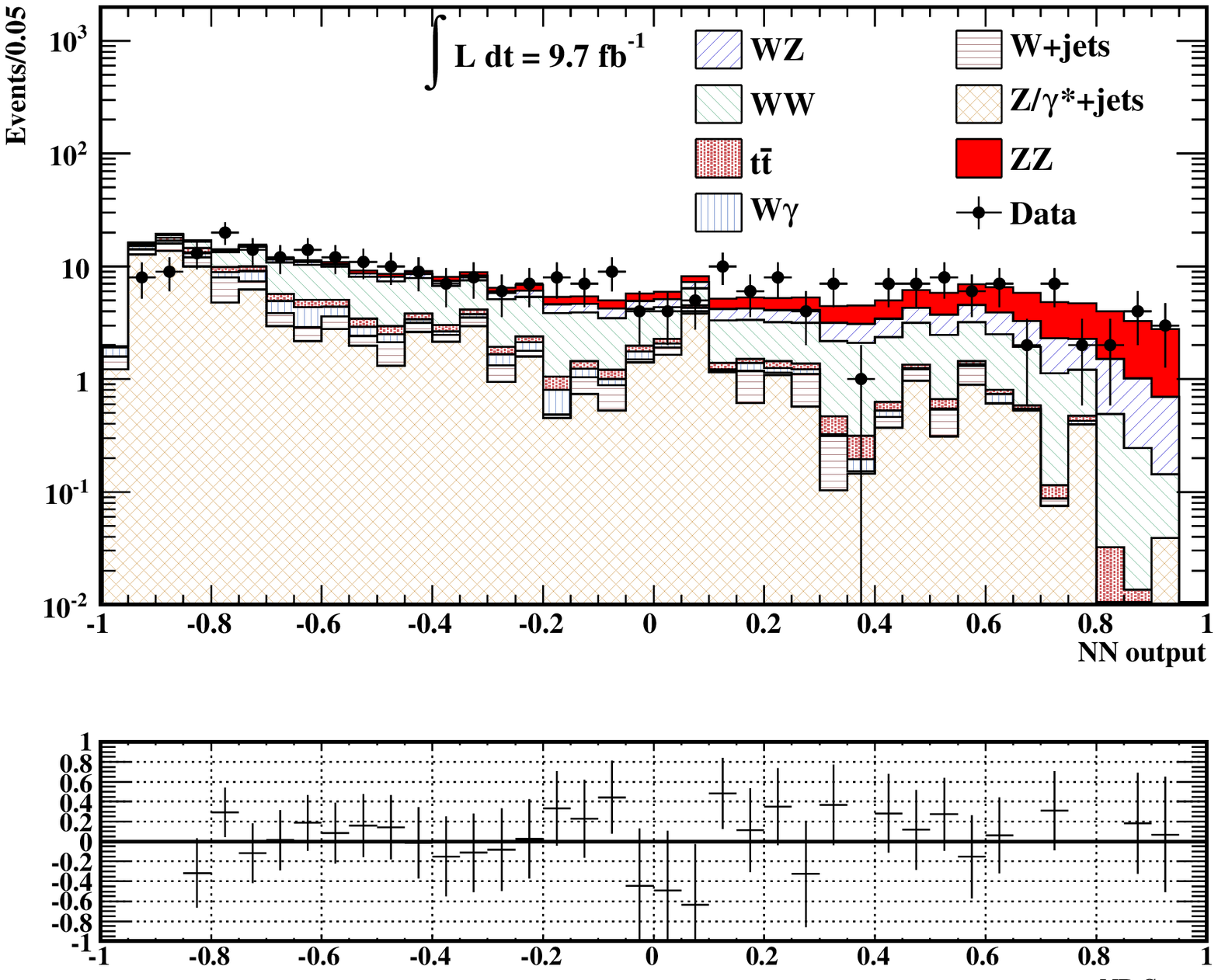}}
\caption{Comparison of predicted and observed NN output distributions 
for the $\llvv$ signal sample shown on (a) linear and (b) logarithmic scales,
including a distribution of the bin-by-bin differences at the bottom.}  
\label{fig:ZZllvv_NN}
\end{figure}

In order to further improve the separation of signal and background, 
we apply an artificial neural network trained on simulated signal and 
background events.  The neural network takes kinematic properties of 
events as inputs and produces a single variable output that is indicative 
of the consistency of the event with either the signal or background 
hypotheses. We use a NeuroBayes neural network (NN)~\cite{NeuroBayes} 
trained on seven kinematic event variables, whose predicted and observed 
distributions for the signal sample are shown in Fig.~\ref{fig:ZZllvv_kin}. 
These variables are, in order of decreasing discrimination power, the 
leading lepton transverse momentum ($p_T(\ell_1)$), the $\MET$ significance 
($\MET/\sqrt{\sum E_T}$), the dilepton invariant mass ($m_{\ell\ell}$), 
the dilepton system transverse momentum ($p_T^{\ell\ell}$), the opening 
angle between the two leptons in the transverse plane ($\Delta\phi(\ell
\ell)$), the number of reconstructed jets ($N_{\rm jets}$), and the angle 
in the transverse plane between the $\MET$ based on energy deposits in 
the calorimeter and a similarly-defined $\MPT$ variable based on charged 
tracks.  A comparison of the predicted and observed distributions of the NN 
output variable for candidate events is shown in Fig.~\ref{fig:ZZllvv_NN}. 
Events consistent with having originated from the signal process are 
assigned NN output values near +1, while those more consistent with having 
originated from one of the background processes have values closer to -1.

\subsection{Systematic uncertainties \label{ssec:llvv_sys}}

The systematic uncertainties considered in this measurement affect 
both predicted signal and background contributions as well as the 
modeled shapes of the NN output variable distribution for each of 
the contributing processes.  Table~\ref{tab:ZZllvv_sys} summarizes 
the complete set of systematic uncertainties incorporated in the 
measurement.

The effect of missing higher-order amplitudes in the simulations 
used to determine detector acceptances is a significant source 
of uncertainty on the predicted event rates for most contributing 
processes.  The sizes of the assigned uncertainties are obtained 
by comparing simulated acceptances from NLO calculations with 
those obtained from the LO event generators.  Uncertainties 
associated with the PDF model taken as input to the simulation are 
assessed following the prescription in Ref.~\cite{Kretzer:2003it}. 
Uncertainties on the theoretical cross sections for 
\WW~\cite{SMXS-old}, \WZ~\cite{SMXS-old}, $W\gamma$~\cite{Baur:1998}, 
and $\ttbar$~\cite{Kidonakis:2003qe,Cacciari:2003fi} production,
which are used to normalize the expected contributions from these
processes, are also incorporated along with the 5.9\% uncertainty 
associated with the CDF luminosity measurement~\cite{Acosta:2002hx}. 
Uncertainties associated with lepton identification and trigger 
efficiency measurements are assessed using the same methodology
described previously for the measurement from the $\llll$ final 
state.  The effect of reconstructed jet energy uncertainties on 
acceptances, including the impact of the veto criteria on events 
containing reconstructed jets, is evaluated in simulation by 
varying the jet energy scale within its measured uncertainties.

\begin{figure}[!htbp]
\centering
\includegraphics[width=\columnwidth]{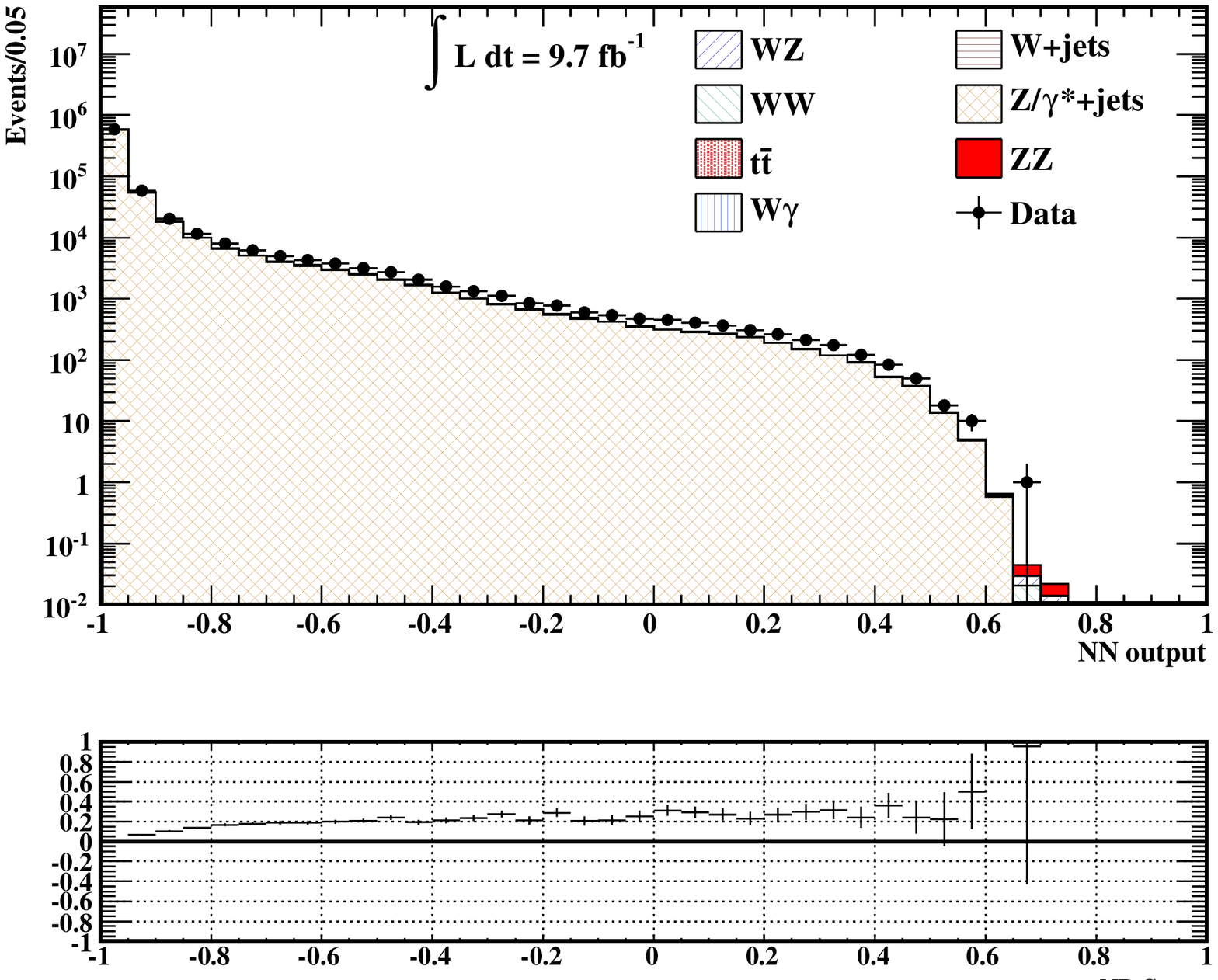}
\caption{Comparison of predicted and observed distributions of 
NN output variable for the low $\MET$ control sample, including 
a distribution of the bin-by-bin differences at the bottom.}
\label{fig:nn_CR_lowMet}
\end{figure}

The uncertainty assigned to the predicted $W$+jets background 
contribution is determined by varying jet misidentification rates 
over the range of values obtained from samples collected with 
different trigger requirements.  The statistical uncertainty 
associated with the fit performed in the high $\MET$ region of 
the $e$--$\mu$ control sample, used to normalize the estimated  
DY contribution, is taken as the systematic uncertainty on the 
event-yield prediction for this process.

\begin{figure}[!htbp]
\centering
\includegraphics[width=\columnwidth]{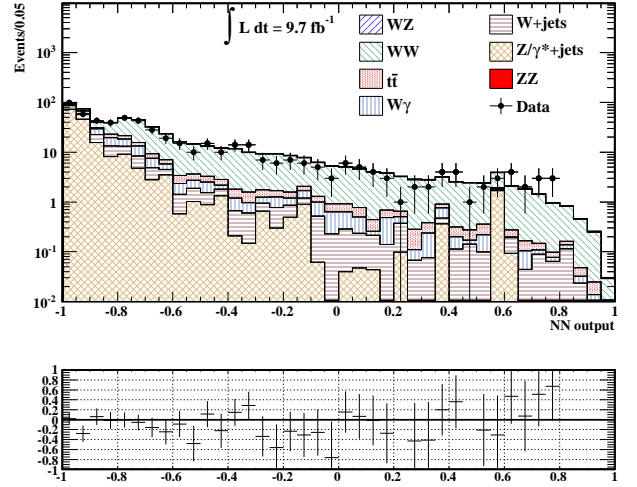}
\caption{Comparison of predicted and observed distributions of 
NN output variable for the $e$--$\mu$ control sample, including 
a distribution of the bin-by-bin differences at the bottom.}
\label{fig:nn_CR_em}
\end{figure}

The mismodeling of relevant kinematic distributions is accounted 
for by incorporating systematic uncertainties covering 
differences between predicted and observed shapes of the NN 
output variable within the previously described control regions.  
Figures~\ref{fig:nn_CR_lowMet} and~\ref{fig:nn_CR_em} show 
comparisons of the predicted and observed distributions of this 
variable in the low $\MET$ and $e$--$\mu$ control samples.  The 
disagreement between predicted and observed distributions in the 
low $\MET$ control sample is used to assess a shape uncertainty 
on the modeled NN output distribution for the DY process, which 
is the dominant contributor of events to this control sample.  
Conversely, due to the agreement between predicted and observed 
distributions of the NN output variable in the $e$--$\mu$ control 
sample, no shape uncertainty is assigned to the modeled NN output 
distribution for the \WW~process, from which a majority of the 
events in this control sample originate.

\begin{table*}[!htbp]
\begin{center}
\caption{Systematic uncertainties incorporated in the cross section 
measurement using the $\llvv$ signal sample. All uncertainties are 
expressed in percents, other than the check mark, representing the shape 
uncertainty considered on the DY simulated prediction.}
\label{tab:ZZllvv_sys}
\begin{ruledtabular}
\begin{tabular}{lccccccc}
Source & \ZZ & \WW & \WZ & $t\overline{t}$ & DY &  $W\gamma$ & $W$+jets \\
\hline
Theoretical cross section            &     & 6   & 6 & 10 &  & 10 & \\
Run-dependence modeling              &     &     &   & 10 &  &  & \\
PDF modeling                         & 2.7 & 1.9 & 2.7 & 2.1 &   & 2.2 & \\
Higher-order amplitudes              & 5   &     & 5 & 10 &  & 5 & \\
Luminosity                           & 5.9 & 5.9 & 5.9 & 5.9 &  & 5.9 & \\
Photon conversion modeling           &     &     &  &  &  & 10 & \\
Jet energy scale                     & 2.0 & 1.6 & 3.4 & 5.3 &  & 2.0 & \\
Jet to lepton misidentification rate &     &     &   &   &   &   &  16 \\
Lepton identification efficiency     & 3   & 3   & 3 & 3 &  &  & \\
Trigger efficiency                   & 2   & 2   & 2 & 2 &  &  &  \\
DY normalization                     &     &     &   &   & 10.2 &  &  \\
DY mismodeling                       &     &     &   &   & $\checkmark$ & & \\
\end{tabular}
\end{ruledtabular}
\end{center}
\end{table*}

\subsection{Result \label{ssec:llvv_result}}

The \ZZ~production cross section is extracted from a fit to the 
NN output variable distribution shown in Fig.~\ref{fig:ZZllvv_NN}.
Following the Bayesian approach used for the measurement in the 
$\llll$ final state, a binned likelihood function is constructed 
from a product of likelihoods for obtaining the results observed 
in each bin based on expected signal acceptance, the number of 
expected background events, and the number of observed events.
Correlations in the signal and background expectations across 
bins are incorporated in the likelihood function as well as 
shared terms for nuisance parameters corresponding to each 
systematic uncertainty source.  The nuisance parameters are 
Gaussian constrained to zero and integrated over their parameter 
spaces in the fit used to extract the cross section measurement.
The value of the cross section that maximizes the constructed 
likelihood, relative to the SM cross section, is $\sigma(\ppbar
\to ZZ)/\sigma^{\rm SM-NLO} =$~0.84~$^{+0.23}_{-0.22}$~(stat)~$^
{+0.16}_{-0.12}$~(syst), which corresponds to a value of 
$\sigma(\ppbar\to ZZ) =$~1.18~$^{+0.32}_{-0.31}$~(stat)~$
^{+0.22}_{-0.17}$~(syst)~pb in the zero-width approximation.

\section{Measurement combination and summary \label{sec:combo}}

We combine the independent \ZZ~production cross section measurements 
obtained from the non-overlapping $\llll$ and $\llvv$ signal 
samples to obtain the final result.  We perform a simultaneous fit, 
considering the number of expected and observed events in the $\llll$ 
sample and the predicted and observed binned NN output variable 
distributions from events in the $\llvv$ sample.  The combination 
procedure takes into account correlations from common systematic 
uncertainty sources affecting signal and background expectations in 
the two samples.  The combined result is
\begin{equation*}
\sigma(\ppbar\to ZZ) = 1.04^{+0.32}_{-0.25}~\mathrm{(stat + syst) ~pb~,}
\end{equation*}
which is consistent with the SM expectation of $\sigma_{ZZ}^{\rm NLO} 
= $~1.4~$\pm$~0.1~pb.  This result, based on the full CDF~II data 
set, approaches the limit in precision achievable at the Tevatron,  
being primarily limited by the size of the available data set.

\section{Acknowledgments}

We thank the Fermilab staff and the technical staffs of the
participating institutions for their vital contributions. This work
was supported by the U.S. Department of Energy and National Science
Foundation; the Italian Istituto Nazionale di Fisica Nucleare; the
Ministry of Education, Culture, Sports, Science and Technology of
Japan; the Natural Sciences and Engineering Research Council of
Canada; the National Science Council of the Republic of China; the
Swiss National Science Foundation; the A.P. Sloan Foundation; the
Bundesministerium f\"ur Bildung und Forschung, Germany; the Korean
World Class University Program, the National Research Foundation of
Korea; the Science and Technology Facilities Council and the Royal
Society, United Kingdom; the Russian Foundation for Basic Research;
the Ministerio de Ciencia e Innovaci\'{o}n, and Programa
Consolider-Ingenio 2010, Spain; the Slovak R\&D Agency; the Academy
of Finland; the Australian Research Council (ARC); and the EU community
Marie Curie Fellowship Contract No. 302103.


\begin{thebibliography}{34}
\bibitem{fakes_foot}
The lepton-like objects used in the
misidentification rate measurement are defined starting from the criteria
applied for lepton identification in Section \ref{sec:event}, relaxing
the requirements on the lepton isolation in the calorimeters and the
fraction of hadronic and electromagnetic energy.
\end{thebibliography}
\end{document}